\documentclass[a4paper,11pt]{article}
\pdfoutput=1 

\usepackage{jheppub} 

\usepackage[T1]{fontenc} 
\usepackage[utf8]{inputenc}
\usepackage{amsmath}
\usepackage{float}
\usepackage{lineno}
\usepackage{xcolor}
\usepackage{subfig}
\usepackage{array}

\usepackage[outdir=source/figures/]{epstopdf}
\usepackage{amssymb}
\usepackage{accents}

\emailAdd{victor.carretero@ific.uv.es}
\emailAdd{tthakore@km3net.de}
\emailAdd{km3net-pc@km3net.de}

\abstract{
In the era of precision measurements of the neutrino oscillation parameters, upcoming neutrino experiments will also be sensitive to physics beyond the Standard Model.  KM3NeT/ORCA is a neutrino detector optimised for measuring atmospheric neutrinos from a few GeV to around 100 GeV. In this paper, the sensitivity of the KM3NeT/ORCA detector to neutrino decay has been explored. A three-flavour neutrino oscillation scenario, where the third neutrino mass state $\nu_3$  decays into an invisible state, e.g. a  sterile neutrino, is considered. We find that KM3NeT/ORCA would be sensitive to invisible neutrino decays with $1/\alpha_3=\tau_3/m_3 < 180$~$\mathrm{ps/eV}$  at $90\%$ confidence level, assuming true normal ordering.
 Finally, the impact of neutrino decay on the precision of KM3NeT/ORCA measurements for $\theta_{23}$, $\Delta m^2_{31}$ and mass ordering have been studied. No significant effect of neutrino decay on the sensitivity to these measurements has been found.
}

\begin{document}

\title{Probing invisible neutrino decay with KM3NeT/ORCA}

\author[a]{S.~Aiello,}
\author[bd,b]{A.~Albert,}
\author[c]{S. Alves Garre,}
\author[d]{Z.~Aly,}
\author[e,f]{A. Ambrosone,}
\author[g]{F.~Ameli,}
\author[h]{M.~Andre,}
\author[i]{M.~Anghinolfi,}
\author[j]{M.~Anguita,}
\author[k]{M. Ardid,}
% ...
\affiliation[a]{INFN, Sezione di Catania, Via Santa Sofia 64, Catania, 95123 Italy}
\affiliation[b]{Universit{\'e}~de~Strasbourg,~CNRS,~IPHC~UMR~7178,~F-67000~Strasbourg,~France}
\affiliation[c]{IFIC - Instituto de F{\'\i}sica Corpuscular (CSIC - Universitat de Val{\`e}ncia), c/Catedr{\'a}tico Jos{\'e} Beltr{\'a}n, 2, 46980 Paterna, Valencia, Spain}
\affiliation[d]{Aix~Marseille~Univ,~CNRS/IN2P3,~CPPM,~Marseille,~France}
\affiliation[e]{INFN, Sezione di Napoli, Complesso Universitario di Monte S. Angelo, Via Cintia ed. G, Napoli, 80126 Italy}
\affiliation[f]{Universit{\`a} di Napoli ``Federico II'', Dip. Scienze Fisiche ``E. Pancini'', Complesso Universitario di Monte S. Angelo, Via Cintia ed. G, Napoli, 80126 Italy}
\affiliation[g]{INFN, Sezione di Roma, Piazzale Aldo Moro 2, Roma, 00185 Italy}
\affiliation[h]{Universitat Polit{\`e}cnica de Catalunya, Laboratori d'Aplicacions Bioac{\'u}stiques, Centre Tecnol{\`o}gic de Vilanova i la Geltr{\'u}, Avda. Rambla Exposici{\'o}, s/n, Vilanova i la Geltr{\'u}, 08800 Spain}
\affiliation[i]{INFN, Sezione di Genova, Via Dodecaneso 33, Genova, 16146 Italy}
\affiliation[j]{University of Granada, Dept.~of Computer Architecture and Technology/CITIC, 18071 Granada, Spain}
\affiliation[k]{Universitat Polit{\`e}cnica de Val{\`e}ncia, Instituto de Investigaci{\'o}n para la Gesti{\'o}n Integrada de las Zonas Costeras, C/ Paranimf, 1, Gandia, 46730 Spain}
% ...
\affiliation[bd]{Universit{\'e} de Haute Alsace, rue des Fr{\`e}res Lumi{\`e}re, 68093 Mulhouse Cedex, France}
% ...

\maketitle
\flushbottom

\section{Introduction}
\label{sec:intro}

The discovery of neutrino oscillations has been one of the early hints of physics beyond the Standard Model (BSM). The three-flavour neutrino model is well established and its current global fit parameters can be found in refs. \cite{Esteban_2019,deSalas:2020pgw}. Even though most of the oscillation parameters are rather well known, several questions remain unanswered, such as the value of the Dirac CP phase $\delta$, the octant of the mixing angle $\theta_{23}$ and the true neutrino mass ordering (NMO), normal ordering (NO) or inverted ordering (IO). Beyond this standard paradigm, the search for potential deviations from the standard scenario, which could arise at sub-leading order, is becoming an important objective for the forthcoming precision neutrino detectors. Such deviations could affect the measurement of the standard neutrino oscillation parameters.

Some of these BSM scenarios are based on the existence of unstable neutrinos. Several theoretical models have been proposed allowing for the decay of neutrinos into a lighter fermion state and a BSM boson. In particular, the Majoron model~\cite{Gelmini:1980re} for Majorana neutrinos, where a neutrino $\nu_i$ can decay into a lighter neutrino $\nu_j$ and a new pseudo-scalar boson, $J$, called Majoron:
 \begin{equation}
     \nu_i \rightarrow \nu_j+J.
 \end{equation}
 
 The Majoron should be dominantly a singlet to comply with the constraints from LEP data on the $Z$ decay to invisible particles \cite{Pakvasa_2000}. On the other hand, if neutrinos are Dirac particles, they could decay from a heavier neutrino to a lighter antineutrino (right-handed singlet) by emitting an iso-singlet scalar boson, $\xi$, carrying lepton number +2 \cite{PhysRevD.45.R1}:
 \begin{equation}
  \nu_i \rightarrow \bar{\nu}_j+\xi.   
 \end{equation}

Neutrinos could decay to visible \cite{visible} or invisible \cite{invisible, Escudero_2020} channels. In the former case, the neutrino created in the decay is an active neutrino and could be detected whereas in the latter, this neutrino is sterile and would remain undetectable. In this paper, we will focus on the invisible neutrino decay independently of the decay model. 

The neutrino decay scenario was proposed as a solution to the solar neutrino problem in 1972 \cite{Bahcall:1972} and as it was shown later, it can only contribute to the solar neutrino deficit at subleading order \cite{Acker_1994}. Stringent bounds on radiative neutrino decay can be set using cosmological data since neutrino mass estimates suggest they would radiatively decay in the microwave energy range \cite{Mirizzi_2007, Chen:2022idm}. Since the three neutrino mass states could have, in principle, non-vanishing masses, $\nu_1$, $\nu_2$ and $\nu_3$ could decay invisibly. However, decays of $\nu_1$ and $\nu_2$ are severely constrained from supernova SN1987A~\cite{Frieman:1987as} and solar neutrino data~\cite{Bandyopadhyay_2003}. The decay of $\nu_3$ can be measured in present accelerator, atmospheric and reactor neutrino experiments. No observation of this phenomenon has been obtained so far, and the best current limits come from combined data analyses such as T2K and NO$\nu$A~\cite{NovaT2K}, MINOS and T2K~\cite{MinosT2K} and K2K, MINOS and SK~\cite{SKK2KMINOS}. Sensitivity studies have been carried out also for future accelerator experiments, like DUNE~\cite{DUNEupdated}, MOMENT~\cite{MOMENT_Tang_2019} and ESSnuSB~\cite{ESSnuSB_choubey2020exploring}, reactor experiments like JUNO~\cite{abrahao2015constraint}, and atmospheric neutrino experiments such as INO~\cite{Choubey_2018_INO}. Astrophysical neutrinos can provide also constraints on neutrino decay. In ref \cite{Denton:2018aml}, the tension observed between track and cascade events in IceCube data is analysed in the context on invisible decay. A more detailed  review on the present and future neutrino decay picture, both invisible and visible, can be found in \cite{Arguelles:2022tki}.

An initial estimate of the  KM3NeT/ORCA sensitivity to invisible neutrino decay was reported in ref. \cite{de_Salas_2019}, showing that bounds from current experiments could be improved by two orders of magnitude. In this paper, we present an updated and more realistic analysis of the KM3NeT/ORCA sensitivity to invisible neutrino decay, where significant improvements on the simulation of the detector response and on the reconstruction performance have been made. In addition, a more complete study of the systematic effects has been carried out.

 This article is organised as follows: in section~\ref{sec:oscframe}, the established scheme of neutrino flavour oscillations both in vacuum and in matter is summarised, as well as the effect that neutrino decay has on the propagation of atmospheric neutrinos. In section~\ref{sec:detector}, the KM3NeT/ORCA neutrino detector is described. In section~\ref{sec:analysis}, the analysis procedure is explained and in section~\ref{sec:results}, neutrino decay sensitivity, potential for discovery, the interplay between the decay parameter, $\alpha_3$, and the mixing angle, $\theta_{23}$, and the effects in KM3NeT/ORCA precision measurements are shown.

\section{Invisible neutrino decay effects in oscillation probabilities}
\label{sec:oscframe}
 
In the standard neutrino oscillation framework, flavour eigenstates,  $\nu_{\beta}$  $(\beta=e, \mu, \tau$) are linearly related to the mass eigenstates, $\nu_i$ $(i=1, 2, 3)$:

\begin{equation}
    \nu_{\beta}=\sum_{i=1}^3 U^*_{\beta i} \nu_i \qquad \implies \qquad \nu_{i}=\sum_{\beta=e,\mu, \tau} U_{\beta i} \nu_{\beta},
\label{Osci1}
\end{equation}

\noindent where $U_{\beta i}$ are the elements of the neutrino mixing matrix, also called Pontecorvo-Maki-Nakagawa-Sakata (PMNS) matrix. This unitary matrix can be parameterised as:

\begin{equation}
    U=\left( {\begin{array}{ccc}
    c_{12}c_{13} & s_{12}c_{13} & s_{13}e^{-i\delta} \\
   -s_{12}c_{23}-c_{12}s_{13}s_{23}e^{i\delta}  & c_{12}c_{23}-s_{12}s_{13}s_{23}e^{i\delta} & c_{13}s_{23} \\
   s_{12}s_{23}-c_{12}s_{13}c_{23}e^{i\delta}  & -c_{12}s_{23}-s_{12}s_{13}c_{23}e^{i\delta}& c_{13}c_{23} \\
  \end{array} } \right),
\end{equation}

\noindent where $s_{ij}$ stands for $\sin\theta_{ij}$ and $c_{ij}$ for $\cos\theta_{ij}$, $\theta_{ij}$ being the three real mixing angles (for a 3-neutrino scenario) and $\delta$ is a  phase accounting for the CP violation.

Neutrino decay can be described by a depletion factor $D=e^{-\frac{t}{\tau_i}}$, where  $\tau_i$ is the rest-frame lifetime of the mass state $m_i$ and $t$ the proper time, rewritten in the lab frame for relativistic neutrinos as $D=e^{-\frac{m_i L}{\tau_i E}}$. This term constitutes the fraction of neutrinos of energy $E$ that survives travelling through a distance $L$. In this work, the invisible neutrino decay is characterised by the parameter $\alpha_i=m_i/ \tau_i$, which is expressed in natural units of $\mathrm{eV^2}$. In order to allow for the invisible neutrino decay, a new term must be included in the Hamiltonian:

\begin{equation}
    H_{\text{Total}}=\frac{1}{2E} \left( H_{0}+H_{M}+H_D \right),
\end{equation}

\noindent where $E$ is the neutrino energy, $H_0$ represents the Hamiltonian in vacuum, $H_M$ represents the coherent scattering on electrons in the neutrino path and $H_D$ accounts for the neutrino decay. After reordering the different terms, the final Hamiltonian is

\begin{equation}
    \centering
   H_{\text{Total}}=\frac{1}{2E} \left[U
  \left( {\begin{array}{ccc}
   0 & 0 & 0 \\
   0  & \Delta m^2_{21} & 0 \\
   0 & 0 & \Delta m^2_{31} \\
  \end{array} } \right)U^{\dagger}+ U
  \left( {\begin{array}{ccc}
   0 & 0 & 0 \\
   0  & 0 & 0 \\
   0 & 0 & -i \alpha_3 \\
  \end{array} } \right)U^{\dagger}  \right]+  
  \left( {\begin{array}{ccc}
   V & 0 & 0 \\
   0  & 0 & 0 \\
   0 & 0 & 0 \\
  \end{array} } \right),
\end{equation}

\noindent with $V=\pm\sqrt{2}N_eG_F$ being the matter potential, $N_e$, the electron density in matter and $G_F$, the Fermi constant. Essentially, the  only  change  in  the  Hamiltonian  is  a shift  in  the  mass  basis term, from $\Delta m_{31}^2$ to $\Delta m_{31}^2-i \alpha_3$.

As a consequence of neutrino decay, the mixing matrix becomes non-hermitian and therefore, the sum of the neutrino oscillation probabilities will differ from unity, due to the disappearance of neutrinos,

\begin{equation}
   P_{\beta e}+P_{\beta \mu}+P_{\beta \tau}=1-P_D(\beta) \qquad  \beta=e,\mu,\tau,
    \label{1minusdecay}
\end{equation}
where $P_D(\beta)$ is the decay probability for flavour $\beta$.

 To study in detail the effects of the decay of the third neutrino mass eigenstate in the oscillatory pattern, the OscProb package~\cite{OscProb} is used.\footnote{OscProb numerical computations are compatible with analytical expressions for neutrino decay in matter derived in refs. \cite{Chattopadhyay:2021eba,Chattopadhyay:2022ftv}.} This package incorporates the possibility to handle invisible neutrino decay in the $\nu_2$ and $\nu_3$ mass eigenstates. To account for matter effects, OscProb uses an Earth density profile model based on radial layers of constant density according to the PReliminary Earth Model (PREM)~\cite{PREM}. In this analysis a total of 44 layers were used. Neutrino decay causes a global decrease of probabilities due to the factor $e^{-\frac{\alpha_3 L}{E}}$ in the non-oscillatory terms, and a damping effect reducing the amplitude of the oscillation terms by a factor
$e^{-\frac{\alpha_3 L}{2E}}$. Therefore, in addition to leading to the observation of fewer neutrinos, the oscillatory pattern will be modified: experiments
with good sensitivity to $\Delta m^2_{31}$  yield a better sensitivity to $\alpha_3$. Furthermore, matter effects play an important role because they can enhance the impact of the invisible decay in the transition probabilities around the resonance region ( $\sim$ $3 - 8$~GeV).  KM3NeT/ORCA will observe an unprecedented number of events in this energy range, allowing to probe matter effects and invisible neutrino decay with precision. 

\begin{figure}
 \centering
  \subfloat{
     \includegraphics[height=5.5cm]{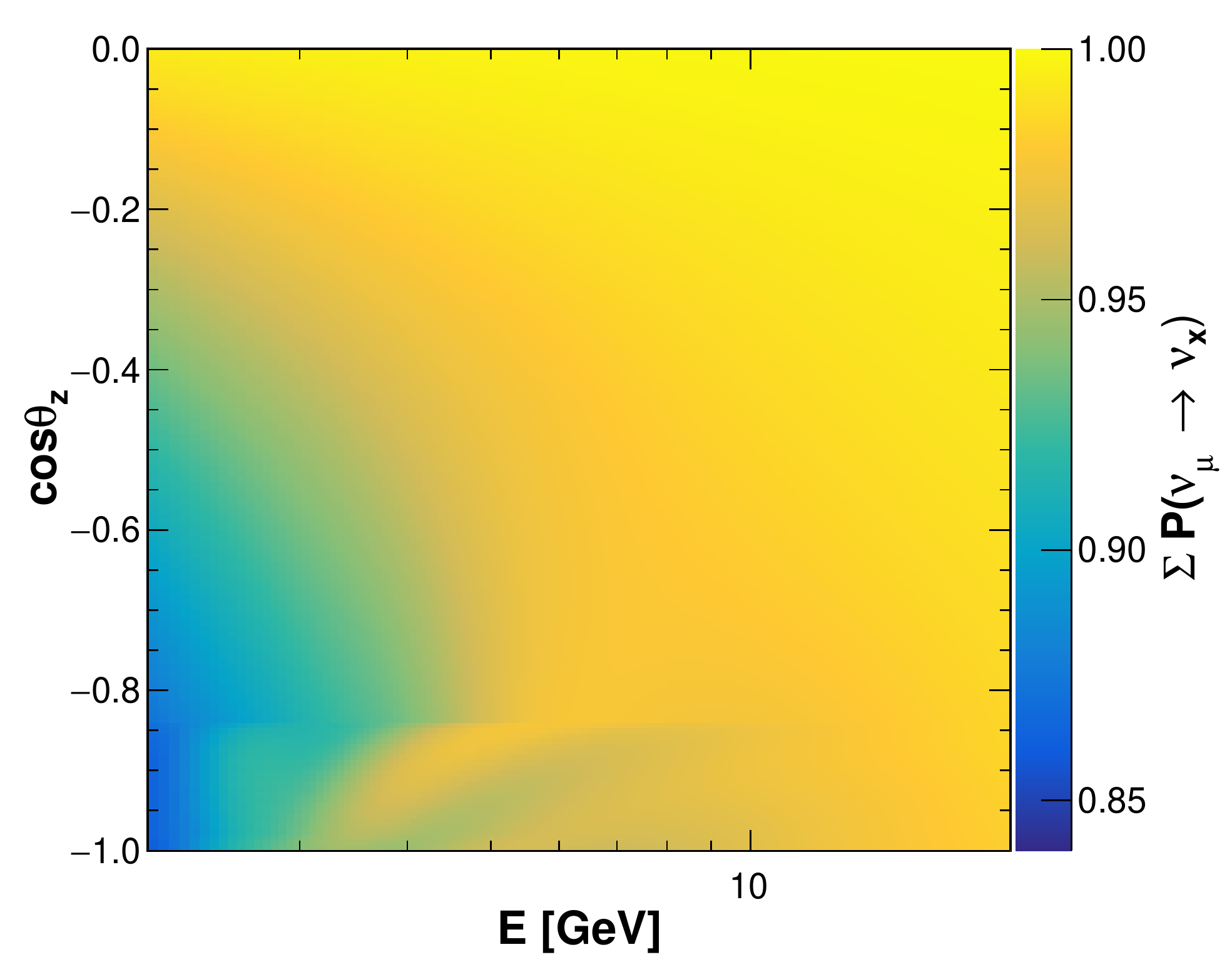}}
  \subfloat{
      \includegraphics[height=5.5cm]{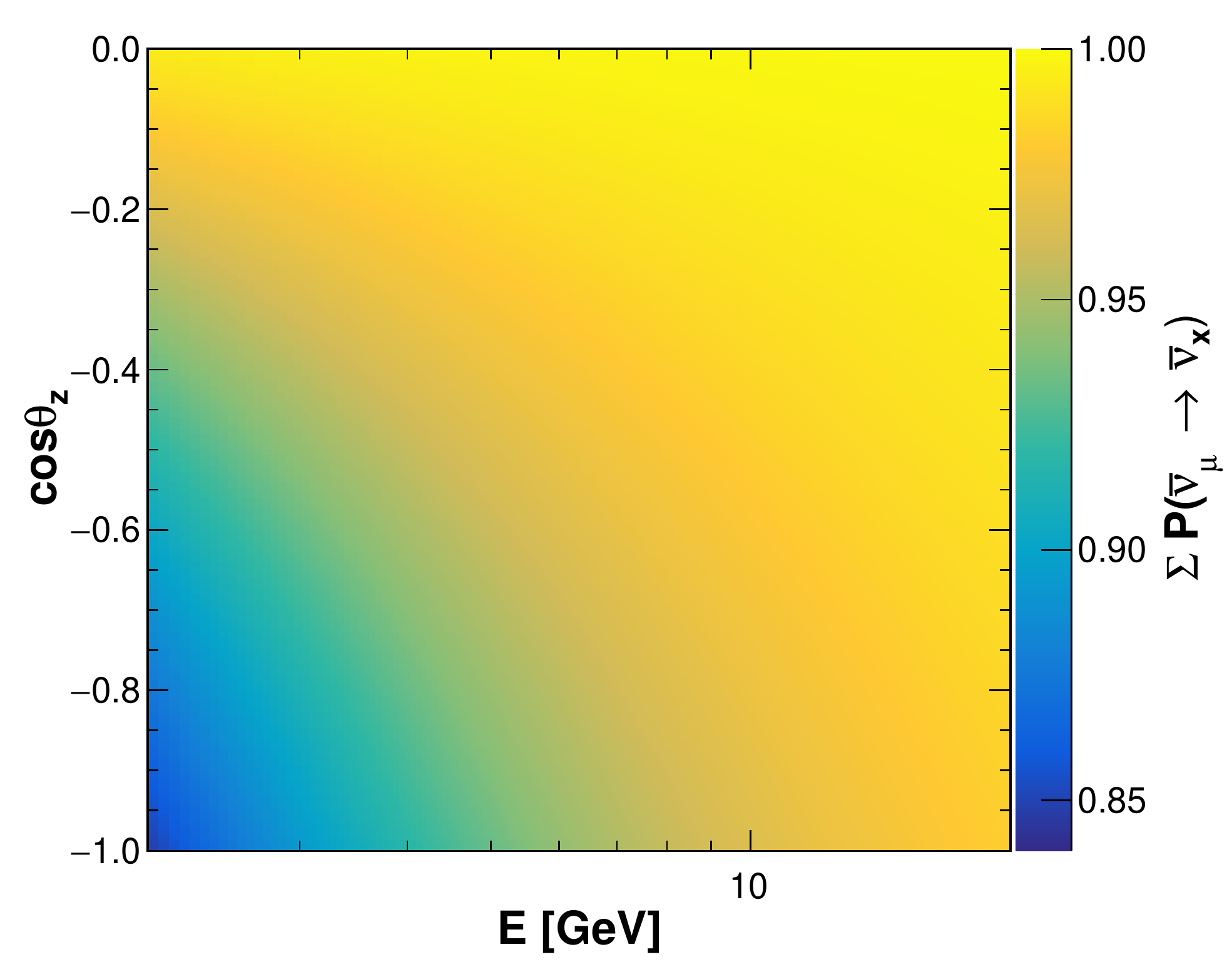}} \\  
\subfloat{
     \includegraphics[height=5.5cm]{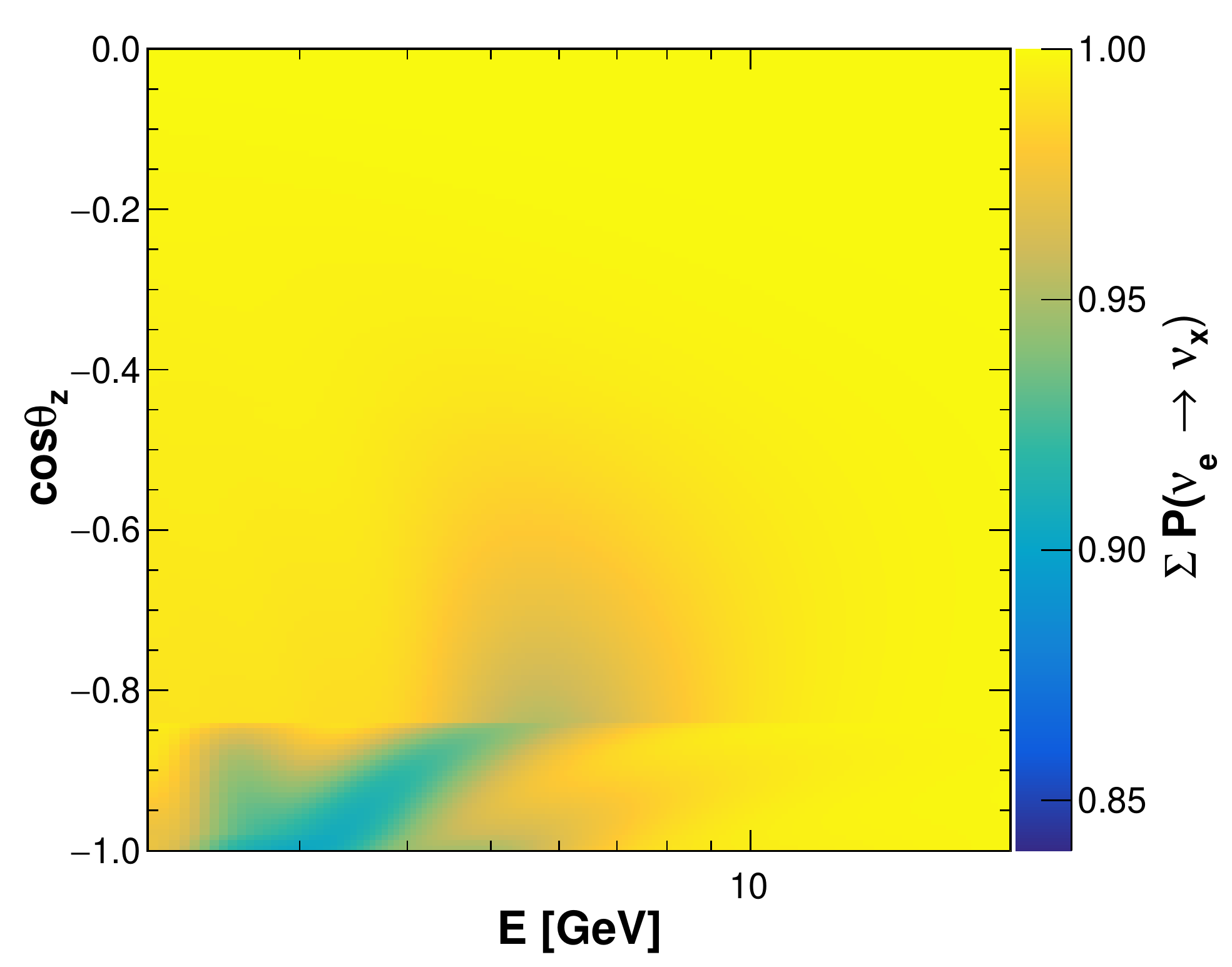}}
  \subfloat{
      \includegraphics[height=5.5cm]{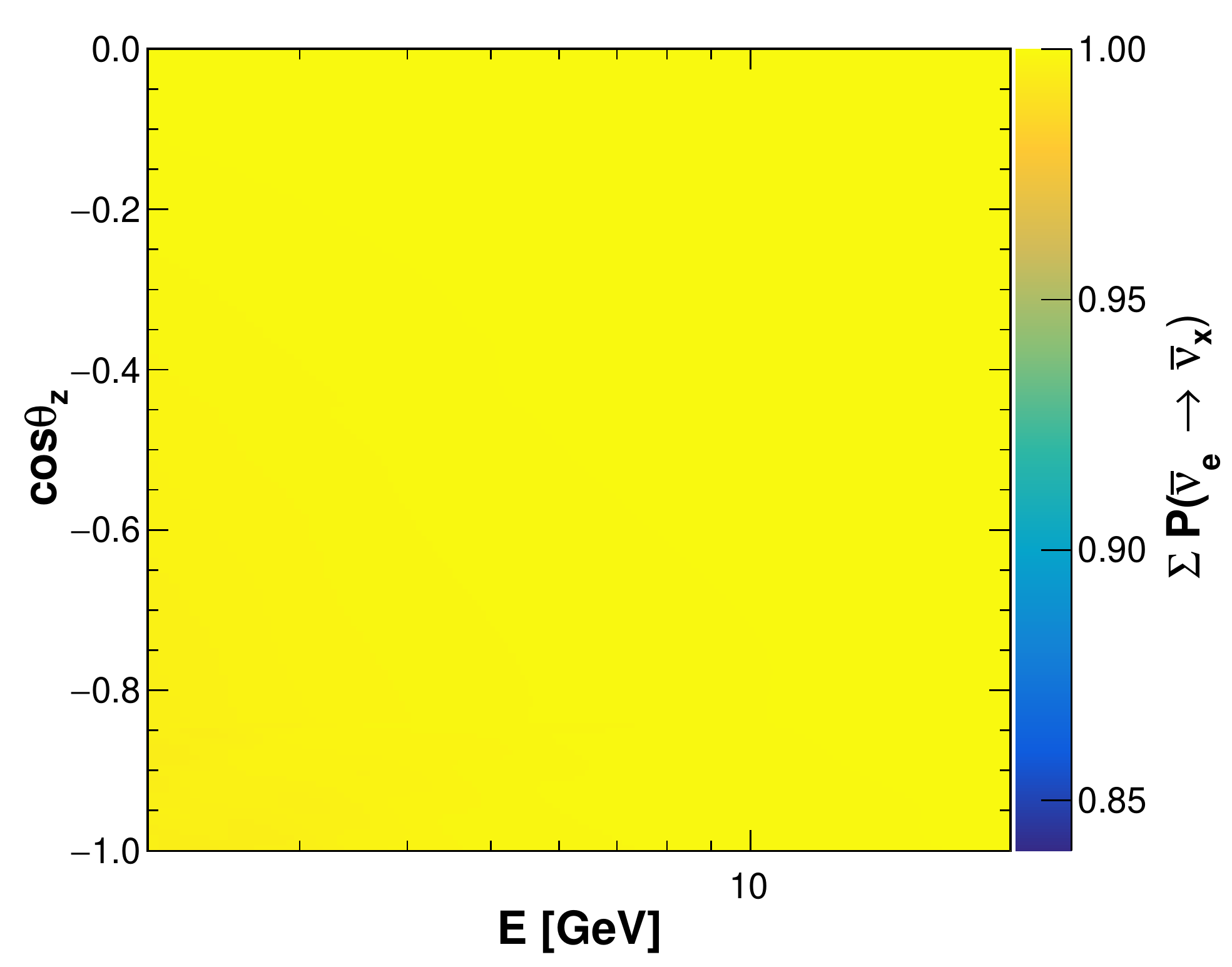}}
 \caption{Sum of muon neutrino (top left), muon antineutrino (top right), electron neutrino (bottom left) and electron antineutrino (bottom right) initial flavour probabilities for $\alpha_3= 10^{-5}~\mathrm{eV^2}$ assuming NO as a function of true neutrino energy and the cosine of the true zenith angle. In the case of IO, probabilities can be described by similar patterns but swapping neutrino and antineutrino. The oscillation parameter values are set to their best-fit values according to NuFit 4.1 \cite{Esteban_2019}.}
\label{Fig1_NC}
 \end{figure}

The breaking of unitarity in eq.~\ref{1minusdecay} for initial muon and electron flavours for both neutrinos and antineutrinos are shown in figure \ref{Fig1_NC}, assuming NO. The oscillation parameter values are taken from NuFit 4.1 \cite{Esteban_2019}.  Although in an experiment like KM3NeT/ORCA it is not possible to distinguish between neutrinos and antineutrinos, both contributions are shown separately. For the muon (anti)neutrino flux, the decrease in the total number of events yields a clear $L/E$ dependence which comes from the depletion factor ($D=e^{-\frac{\alpha_3 L}{ E}}$). In the case of the electron (anti)neutrino flux, the behaviour is different due to the small value of $\theta_{13}$, making $\nu_3$ contributions to the electron flavour around 2\% in the absence of matter effects. Antineutrino probabilities are almost non-affected by decay whereas neutrino probabilities are significantly damped around the matter-resonance region due to the enhancement of the electron neutrino transition probability, $P_{e\mu}$, suppressed by decay in this case.

\begin{figure}
 \centering
  \subfloat{
   \label{Fig2_Osc-1}
     \includegraphics[height=4.3cm]{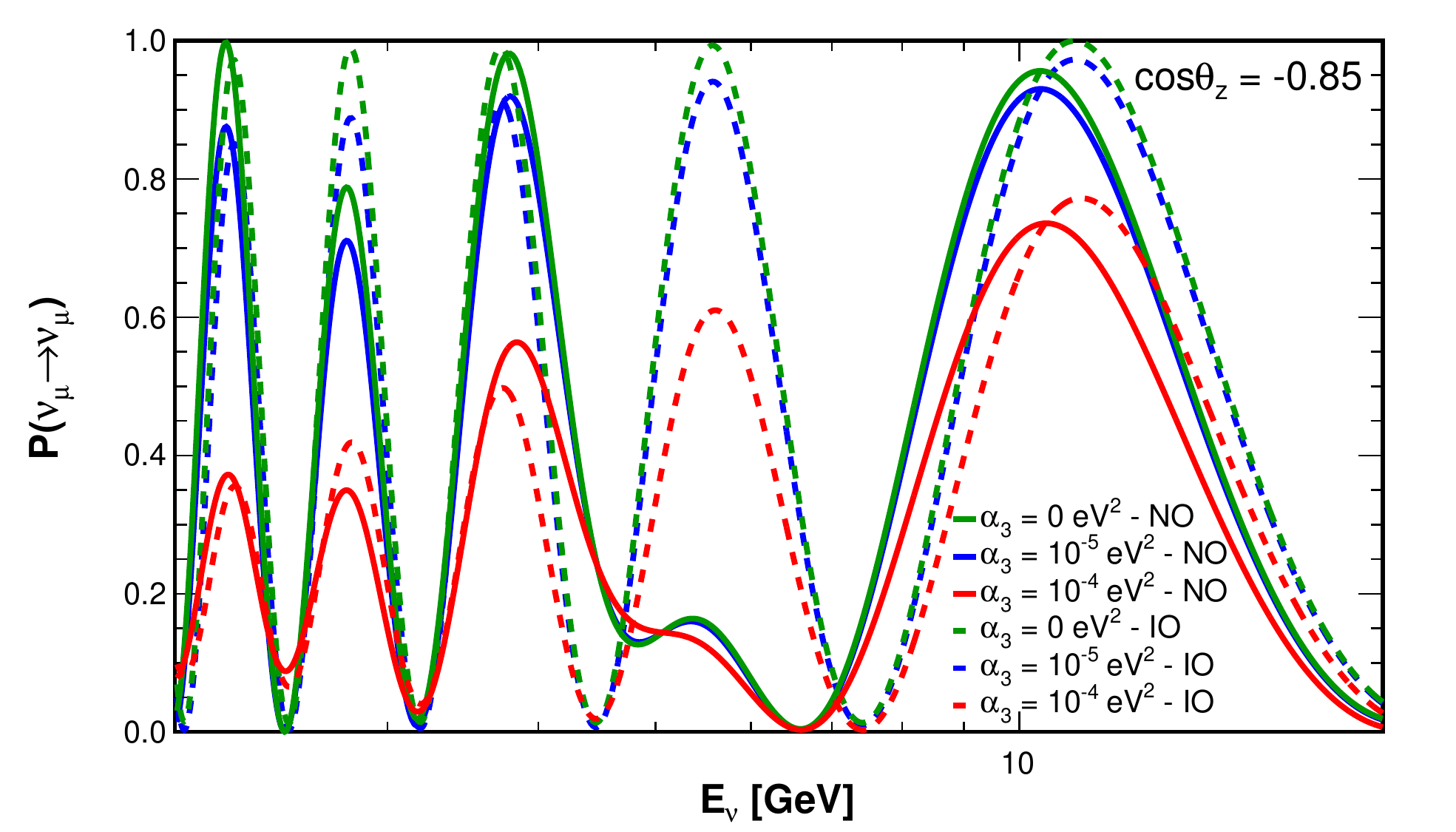}}
  \subfloat{
   \label{Fig2_Osc-2}
      \includegraphics[height=4.3cm]{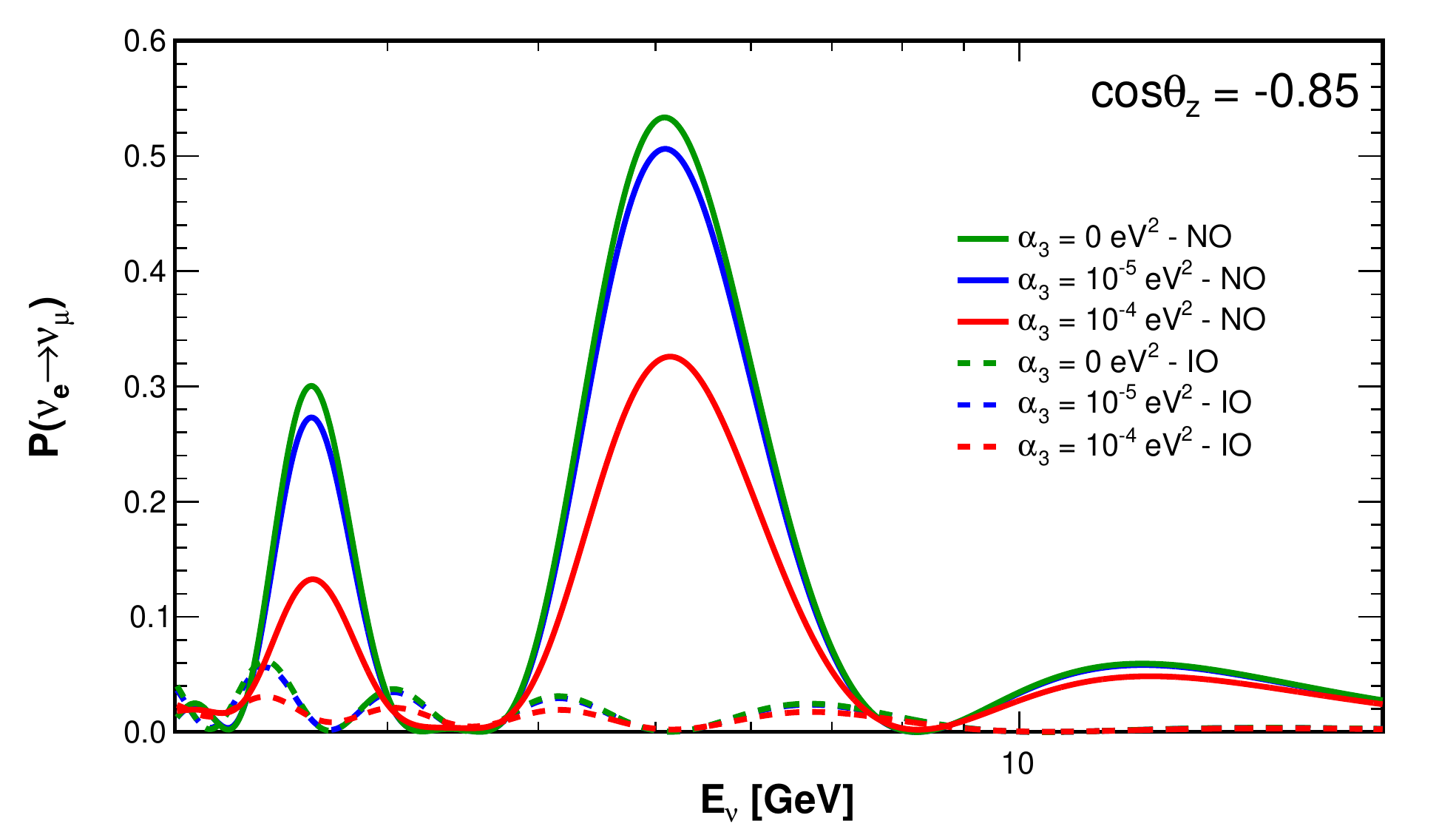}} \\
       \subfloat{
   \label{Fig2_Osc-3}
     \includegraphics[height=4.3cm]{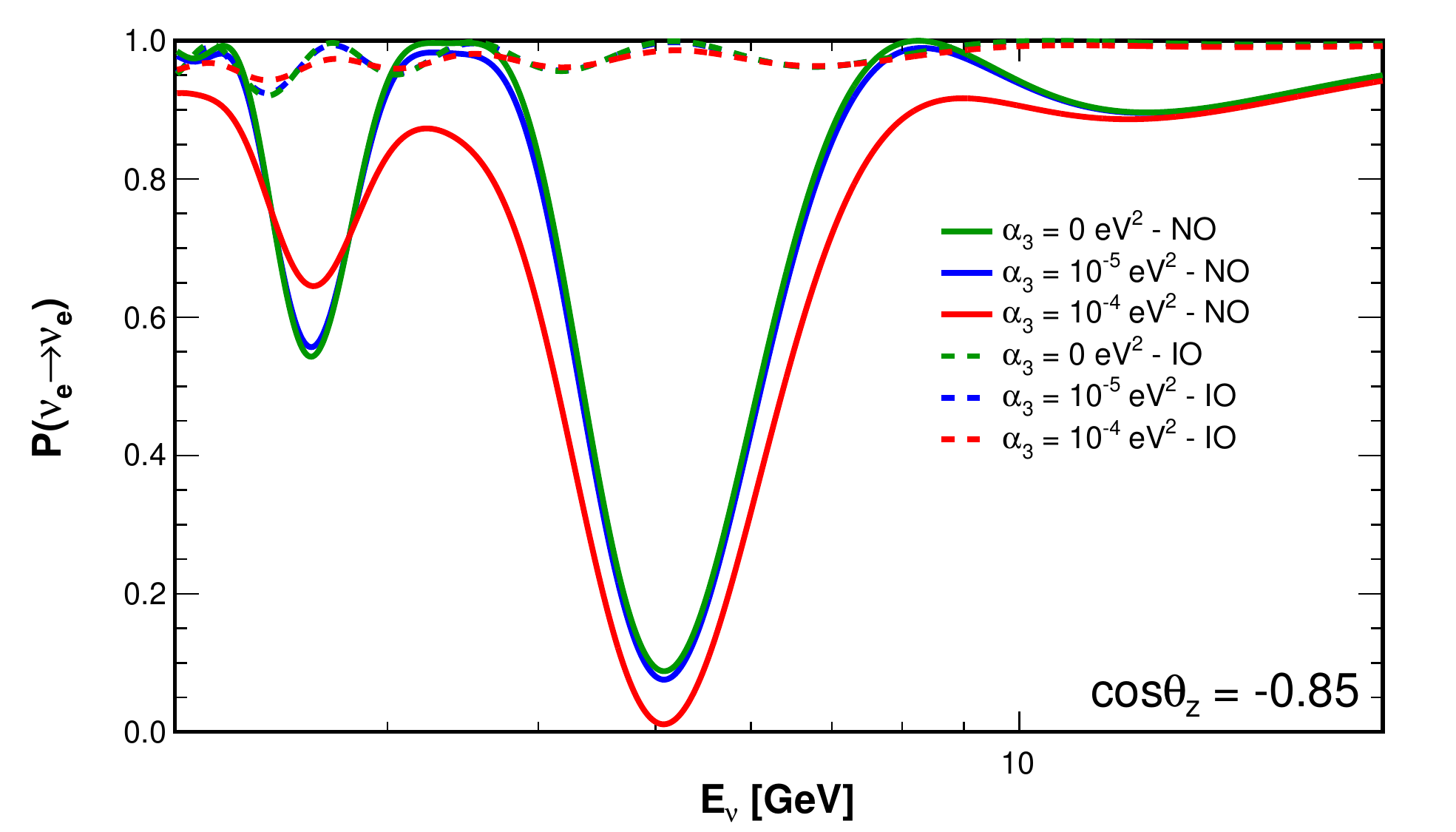}}
  \subfloat{
   \label{Fig2_Osc-4}
      \includegraphics[height=4.3cm]{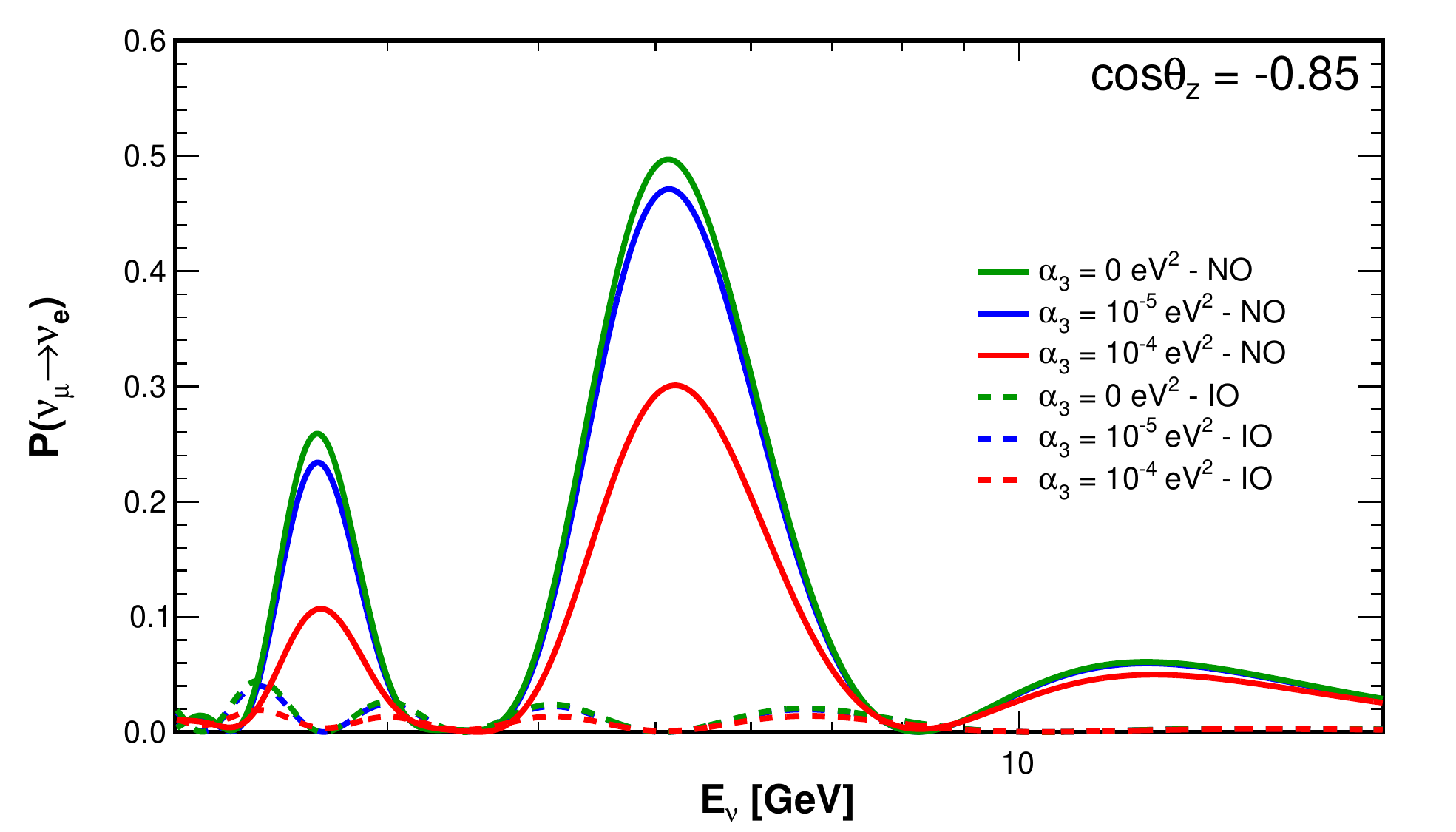}}

 \caption{Muon neutrino survival (top left), transition to muon neutrino (top right), electron neutrino survival (bottom left) and transition to electron neutrino (bottom right) probabilities as a function of energy at a cosine of the zenith angle $\cos \theta_z = - 0.85$. Three values of the decay constant are  considered:  $\alpha_3 = 0$ (green),  $\alpha_3 = 10^{-5}~\mathrm{eV^2}$ (blue) and $\alpha_3 =  10^{-4}~\mathrm{eV^2}$ (red). The solid (dashed)  curves are for NO (IO). Antineutrino probabilities can be described by the same curves but swapping the orderings.  }
 \label{Osc3}
\end{figure}

 The damping effects of neutrino decay in the oscillation pattern can be observed in figure \ref{Osc3}, which shows decay effects in the muon (electron) neutrino survival (transition) probability. Decay effects induced by $\alpha_3$ are stronger in channels related to the muon flavour because $\nu_{\mu}$ has larger content of $\nu_3$ than $\nu_e$. In fact, in the absence of matter effects, the channel  most affected by neutrino decay is $P_{\mu\mu}$, regardless of the mass ordering.
However, the matter enhancement of the transition probabilities $\nu_{\mu} \leftrightarrow \nu_{e}$ if normal ordering is true and $\bar{\nu}_{\mu} \leftrightarrow \bar{\nu}_{e}$ if inverted ordering is true, makes these channels more sensitive to neutrino decay in the energy region around the resonance. The expected effects are similar in both orderings, but taking into account that the neutrino-nucleon cross section is approximately twice that for antineutrinos and that the atmospheric neutrino flux is ~1.1 times the antineutrino flux, the KM3NeT/ORCA sensitivity to neutrino decay will be lower in the case of IO.

\begin{figure}[H]
 \centering
  \subfloat{
   \label{Fig3_Cor1}
     \includegraphics[height=4.3cm]{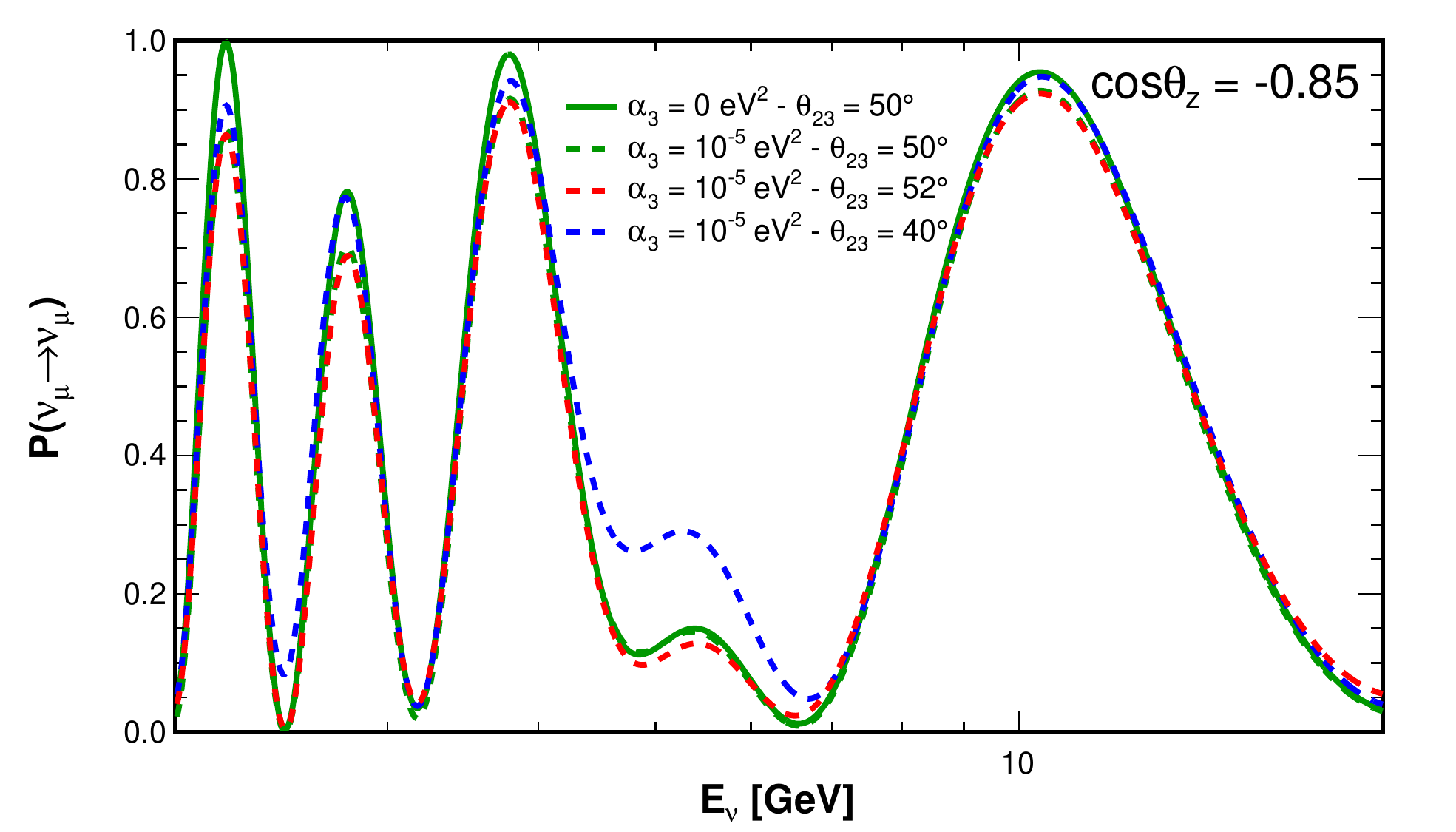}}
  \subfloat{
   \label{Fig3_Cor2}
      \includegraphics[height=4.3cm]{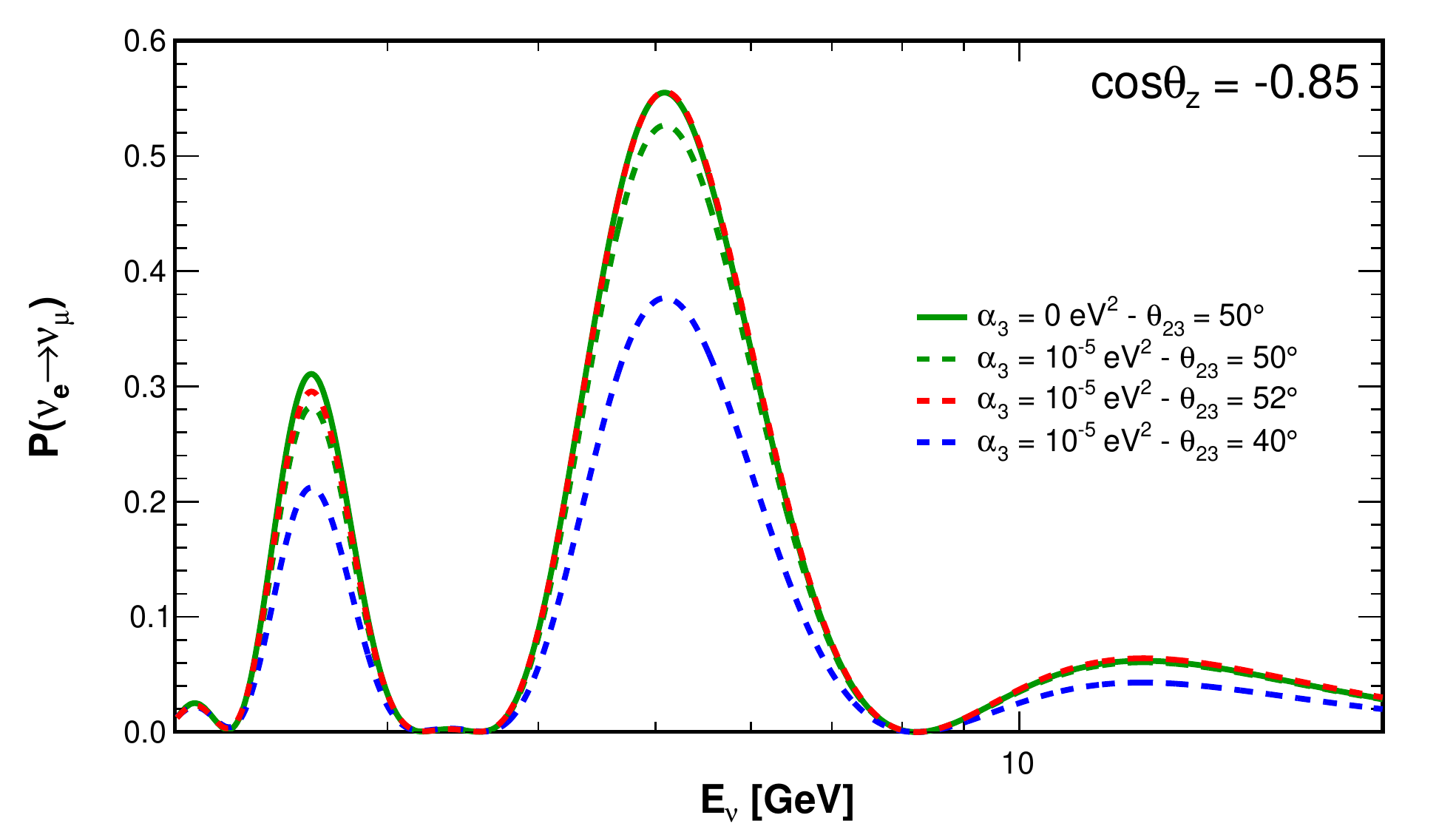}}

 \caption{Muon neutrino survival (left) and electron neutrino oscillation to muon neutrino (right) probabilities as a function of energy at a cosine of the zenith angle $\cos \theta_z = - 0.85$ assuming NO. Four cases are shown: $\alpha_3 = 0$ with $\theta_{23}=50^{\circ}$ (solid green), $\alpha_3=10^{-5}~\mathrm{eV^2}$ with the same value of $\theta_{23}$ (dashed green) and with two different values: $\theta_{23}=52^{\circ}$ (dashed red) and $\theta_{23}=40^{\circ}$ (dashed blue). }
 \label{Th23Change}
 \end{figure}

  There is a subtle correlation between $\alpha_3$ and $\theta_{23}$ with a different behaviour for the oscillation and survival channels. The interplay between $\theta_{23}$ and $\alpha_3$ has been extensively studied in refs. \cite{ESSnuSB_choubey2020exploring, FirstSecondMin} for specific baselines. For atmospheric neutrino experiments with a broader range of baselines this correlation is more complicated.
Figure \ref{Th23Change} shows the muon neutrino survival (electron neutrino oscillation to muon) probability for four cases of $\theta_{23}-\alpha_3$ values. The decrease of the survival probability $P_{\mu\mu}$ at the oscillation maxima due to decay effects could be partially compensated by reducing the value of $\theta_{23}$ to the lower octant, but this would increase the probability in the energy range where matter effects are relevant. However, in the case of the transition probability $P_{e\mu}$, a higher value of $\theta_{23}$ compensates the decrease caused by the decay effects. These anti-correlations will play a role when constraining the invisible decay parameter and will be discussed in section \ref{sec:th23}.

\section{The KM3NeT/ORCA detector}
\label{sec:detector}
KM3NeT \cite{Adri_n_Mart_nez_2016} is a research infrastructure hosting a network of neutrino experiments placed on the seabed of the Mediterranean Sea. The infrastructure is currently under construction and it is distributed in two sites. The two detectors, using the same technology, pursue different physics objectives according to their different spatial configuration. The KM3NeT/ARCA detector (Astroparticle Research with Cosmics in the Abyss) is being installed in the KM3NeT-It site, 100 km offshore the Sicilian coast near Capo Passero (Italy) at a depth of 3500~m. KM3NeT/ARCA searches for high energy neutrinos from astrophysical sources. The KM3NeT/ORCA detector (Oscillation Research with Cosmics in the Abyss) is being built near the coast of Toulon (France), 40~km offshore and at 2500~m depth. KM3NeT/ORCA exploits the neutrino flux generated in the Earth's atmosphere and the matter effects experienced when neutrinos traverse the Earth to determine the neutrino mass ordering \cite{NMOpaper}. Using atmospheric neutrinos, KM3NeT/ORCA will study BSM effects in the neutrino sector including non-standard interactions \cite{KhanChowdhury:2021kce}, sterile neutrinos \cite{sterile} and the topic of this paper, invisible neutrino decay.
 
Arrays of thousands of optical sensors detect the Cherenkov photons induced by charged particles originating from neutrino interactions.  The optical photosensors are hosted in pressure-resistant glass spheres called Digital Optical Modules (DOMs). Each DOM \cite{DOMPAPER} houses 31 photomultiplier tubes (PMTs) and their associated readout electronics. The DOMs are arranged along vertical flexible string structures called Detection Units (DUs) which are anchored to the sea floor and kept vertical by the buoyancy of the DOMs and by a submerged buoy at the top.  Each DU contains 18 DOMs organised along two parallel thin ropes. A collection of 115 DUs forms a building block. KM3NeT/ARCA will be formed by two blocks and KM3NeT/ORCA will be formed by one single block. The average horizontal spacing between the strings for KM3NeT/ARCA is $\sim90$~m and for KM3NeT/ORCA is $\sim20$~m. The vertical spacing between the DOMs along the string for KM3NeT/ARCA is $\sim36$~m and for KM3NeT/ORCA is $\sim9$~m. The instrumented volume is around 1~$\mathrm{km^3}$ for KM3NeT/ARCA and $7\times 10^6~\mathrm{m^3}$ for KM3NeT/ORCA. The spacing between DOMs affects the energy threshold, making KM3NeT/ARCA more suitable for studies in the TeV-PeV energy range and KM3NeT/ORCA for the GeV energy range.

\section{Analysis}
\label{sec:analysis}

The detector response to the neutrino flux is simulated with Monte Carlo (MC) events as described in \cite{NMOpaper}. Neutrino events in KM3NeT/ORCA are generated by gSeaGen \cite{Aiello:2020}, a GENIE-based \cite{andreopoulos2015genie} application developed to efficiently generate high statistics samples of events induced by neutrino interactions and detectable in a neutrino telescope. The particles produced in neutrino interactions are propagated through the sea water to the instrumented volume with KM3Sim \cite{Tsirigotis:2011zza}, a package based on GEANT4 \cite{AGOSTINELLI2003250}. Cherenkov photons induced by primary and secondary particles are also propagated to the PMTs by KM3Sim, taking into account the light absorption and scattering in sea water. To generate atmospheric muon events, MUPAGE \cite{Carminati:2008qb, parametrization} provides parameterised muon bundles at the detector surface, and the KM3 package \cite{MCAntares} tracks them in sea water producing the subsequent Cherenkov photons. The optical backgrounds due to the PMT dark count rate and due to the decays of $^{40}\mathrm{K}$ present in sea water are included through a custom-KM3NeT package, which also simulates the digitised output of PMT responses and the readouts. The atmospheric muon background is reduced by selecting only upgoing events, since atmospheric muons cannot traverse the Earth. More details of simulations as well as the improvements in terms of triggering with respect to the LoI \cite{Adri_n_Mart_nez_2016} can be found in ref. \cite{NMOpaper}.

 At the energies most relevant to KM3NeT, neutrino interactions in sea water are mainly events of deep inelastic scattering between neutrino and nucleons, though quasi-elastic scattering cannot be neglected.  Charged-current (CC) neutrino interactions produce a hadronic shower and a charged lepton. Neutral-current (NC) neutrino interactions produce a neutrino and a hadronic shower. Muons at GeV energies experience a roughly constant energy loss, travelling in approximately straight lines and giving rise to a long track of uniform brightness. Electrons and hadrons develop  particle showers with typical lengths of a few meters, appearing as localised light sources in the detector in contrast to the long muon tracks that span tens to hundreds of meters. Consequently, events in KM3NeT/ORCA are classified in two topologies: $track$-$like$ events ($\nu_{\mu}$~CC events and $\nu_{\tau}$~CC events in which the produced tau decays into a muon), and $shower$-$like$ events ($\nu_{e}$~CC events, the rest of the decay channels of $\nu_{\tau}$~CC events and all $\nu$ NC events). The same topologies apply for antineutrinos. The energy and direction of the events are reconstructed assuming a given topology (track-like or shower-like) and applying a maximum likelihood fit.

 To classify events into these topologies, three machine learning random decision forest (RDFs) algorithms are used. The first two are designed to  suppress background events (one for optical noise and one for atmospheric muons) and the third one is optimised to distinguish between the track or shower topologies.  Each trained classifier yields a score variable (atmospheric muon score, noise score, track score) and events are selected or rejected according to cuts in terms of the RDF scores. Although two topologies are identified and reconstructed, three event classes are defined for the analysis: shower events, intermediate events and track events.  Events with a track score lower than $0.3$ are considered shower-like events. If the track score is above $0.7$ they are classified as track-like events. The rest of the events in between these two cuts are assigned to the intermediate class. Background is suppressed by rejecting events with a muon score greater than $0.05$ and a noise score bigger than $0.1$. A more detailed description of the selection, reconstruction and classification of the events can be found in ref. \cite{NMOpaper}.

The analysis is based on the expected 2-dimensional distributions in reconstructed energy and reconstructed cosine of the zenith angle for each of the three event classes (tracks, intermediates and showers). These distributions are obtained with a KM3NeT analysis framework, called SWIM \cite{Bourret:2018kug},  developed to model the detector response using MC simulations. These distributions are obtained by first calculating the true energy and zenith angle distributions for each (anti)neutrino interaction type.~Afterwards, the interaction rates at the detector are computed as described in ref. \cite{Bourret:2018kug} taking into account cross sections, neutrino fluxes, the effective mass and the neutrino oscillation probabilities. 

The detector resolution is taken into account by means of a response matrix which is evaluated by reconstructing MC events. This matrix is used to relate the true and the reconstructed variables used in the analysis. For each interaction channel $\nu_x$ and each classification $i$, a 4-dimensional response matrix $R^{[\nu_x\rightarrow i]}(E_{\text{true}},\theta_{\text{true}},E_{\text{reco}},\theta_{\text{reco}})$ is defined. Each entry represents the efficiency of detection, classification and probability of reconstruction for a given true bin $(E_{\text{true}},\theta_{\text{true}})$.

The binning and ranges used for this analysis are listed in table \ref{Table1-binning}. Energy bins are equally spaced in  logarithmic scale and $\cos\theta$ bins are uniform in linear scale. The ‘Beeston and Barlow light method’ \cite{BARLOW1993219} was used to account for statistical uncertainties due to finite MC statistics. A sufficient number of events in each bin was guaranteed to assure a smooth sampling of the detector response. 

\begin{figure}[H]
 \centering
  \subfloat{
   \label{Fig4_Map_T}
     \includegraphics[height=3.9cm]{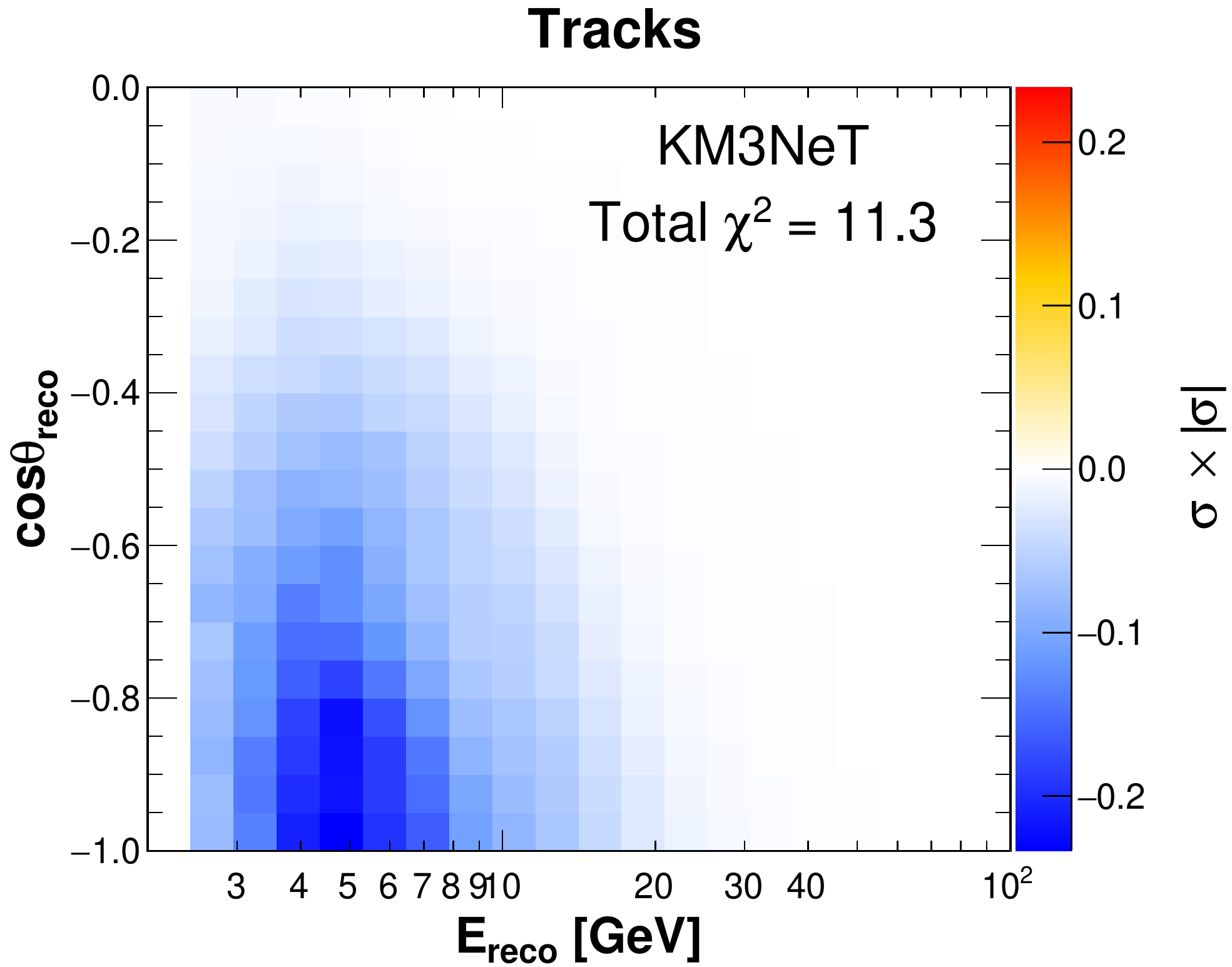}}
  \subfloat{
   \label{Fig4_Map_I}
     \includegraphics[height=3.9cm]{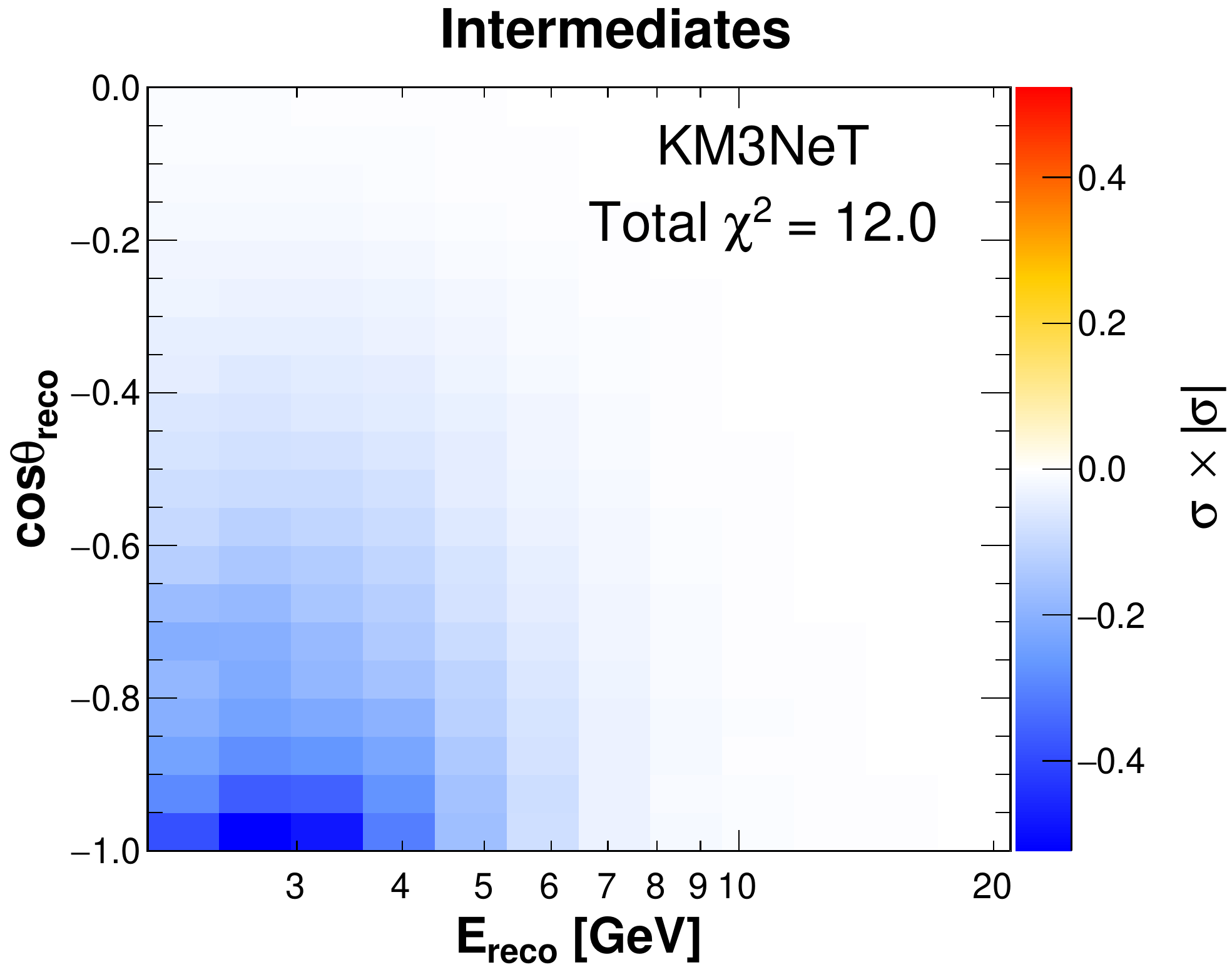}}
 \subfloat{
   \label{Fig4_Map_S}
     \includegraphics[height=3.9cm]{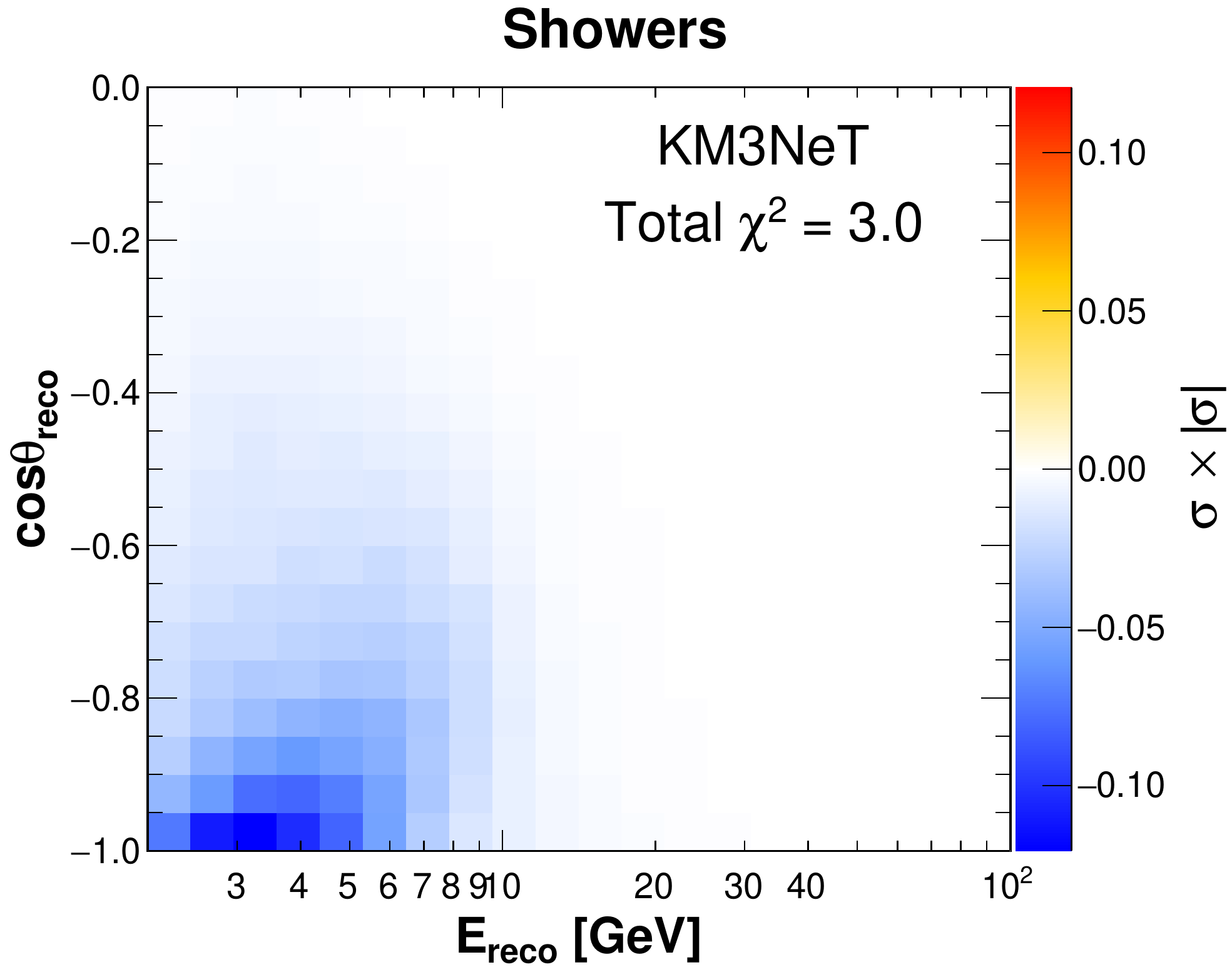}}
 
 \caption{Statistical signed-$\chi^2$ distributions comparing standard oscillations to neutrino decay with $\alpha_3=10^{-5}~\mathrm{eV^{2}}$ for 1 year of data taking in KM3NeT/ORCA assuming NO as a function of reconstructed energy and reconstructed cosine of the zenith angle for track events (left), intermediate events (middle) and shower events (right). Note the different scaling.}
 \label{signedchi2}
\end{figure}

\begin{table}[H]
    \centering
 \begin{tabular}{|c|c|c|c|c|}
 \hline
     & $E_{\text{true}}$[GeV] & $\cos\theta_{\text{true}}$ & $E_{\text{reco}}$[GeV]  & $\cos\theta_{\text{reco}}$\\ \hline
    Bins & 40 & 80 &20 &20  \\ \hline
    Range &[1, 100] & [$-$1, 1] & [2, 100] & [$-$1, 0] \\ \hline
\end{tabular}
    \caption{Bin choice for the MC-based response matrix used in this analysis. Energy bins are equally spaced in
$\log_{10}$.}
    \label{Table1-binning}
\end{table}
In order to illustrate the influence of neutrino decay on the reconstructed energy – cosine zenith angle distributions, a statistical signed-$\chi^2$ is computed for $\alpha_3=10^{-5}~\mathrm{eV^2}$ and 1 year of data taking. The result is shown in figure \ref{signedchi2}. The signed-$\chi^{2}$ is defined as:
 \begin{equation}
     \chi^2_S=\sigma \cdot |\sigma|;~\sigma=  \frac{N_{\text{D}}-N_{\text{STD}} }{\sqrt{N_{\text{STD}}}},
     \label{Eq:statchi2}
 \end{equation}
 where $N_\text{D}$ and $N_{\text{STD}}$ are the expected number of events in the neutrino decay and in the standard scenarios, respectively,  as a function of the reconstructed energy
and zenith angle. The regions in this phase space that contribute the most to neutrino decay coincide with what would be expected due to the depletion factor of the form: $D=e^{-\frac{\alpha_3 L}{ E}}$, i.e., low energies and longer neutrino paths. Since particle identification is difficult at very low energies, most of the $\nu_{\mu}$~CC events between 2 and 4~GeV are classified as intermediate events, thus increasing the contribution of this sub-sample to the sensitivity.
 \clearpage
The analysis procedure to constrain the invisible decay parameter $\alpha_3$  is based on the maximisation of a binned log-likelihood of the 2-dimensional distribution of events in log$_{10}$($E_{\text{reco}}$/GeV) and $\cos \theta_{\text{reco}}$ comparing the observed data to a model prediction, assuming a particular value of $\alpha_3$. The sensitivities are computed using the Asimov approach, where the observed data is replaced by a representative dataset defined as the one  which provide the expected values of the null hypothesis in each bin  \cite{Cowan:2010js}. The representative Asimov dataset is computed assuming standard oscillation parameters from NuFit 4.1 \cite{Esteban_2019}. The log-likelihood is modelled as a combination of Poisson distributions for the expected number of events in each bin and Gaussian distributions associated with the nuisance parameters. Following Wilks' theorem \cite{Wilks:1938dza}, the negative log-likelihood behaves as a chi-square distribution:

\begin{equation}
    \chi^2=-2 \log L =\chi^2_{\text{stat}}+\chi^2_{\text{syst}},
\end{equation}

\begin{align}
     \chi^2(\alpha_3)= \underset{{\vec{\epsilon}}}{\mathrm{min}}   \left \{ 2 \sum_{i,j}  \Biggl[ (N^{\text{mod}}_{ij}(\alpha_3 ; \vec{\epsilon})-N^{\text{dat}}_{ij})+N^{\text{dat}}_{ij} \log \left( \frac{N^{\text{dat}}_{ij}}{N^{\text{mod}}_{ij}(\alpha_3;\vec{\epsilon})}\right) \Biggr] \right .  + \\ \left . +\sum_k \left(\frac{\epsilon_k-\langle\epsilon_k\rangle}{\sigma_k}\right)^2 \right \} . \nonumber
    \label{totalchi}
\end{align}

 $N^{\text{mod}}_{ij}$ and $N_{ij}^{\text{dat}}$ represent the number of reconstructed events expected by the model and the number of events observed, respectively, in the bin ($i$,$j$). The parameters of the model that characterise the distributions ($\vec{\epsilon}$) are composed by oscillations parameters (in table \ref{Table2-Nufit}), and nuisance parameters, which are related to systematic uncertainties (in table \ref{Table3-Nuisance}).  Some of these parameters are constrained with priors representing constraints from other experiments (mean value $\langle\epsilon_k\rangle$ and standard deviation $\sigma_k$). Specifically:
\begin{enumerate}
    \item The normalisation of the three classes is allowed to vary without constraints.
    \item The normalisation of NC events is allowed to vary without constraints.
    \item The ratio skew between the total number of $\nu_{\mu}$ and $\bar{\nu}_{\mu}$, introduced as $1+r_{\mu\bar{\mu}}$, varies with a standard deviation of 5\% of the parameter's nominal value.
    \item The ratio skew between the total number of $\nu_{e}$ and $\bar{\nu}_{e}$ varies, introduced as $1+r_{e\bar{e}}$, with a standard deviation of 7\% of the parameter's nominal value.
    \item The ratio skew between the total number of $\nu_{\mu}$ and ${\nu_{e}}$, introduced as $1+r_{\mu e}$, varies with a standard deviation of 2\% of the parameter's nominal value.
     \item The ratio skew of upgoing to horizontally-going neutrinos, denoted as the zenith slope, varies introducing a tilt as, $1+\epsilon_{\theta}\cos\theta$, with a standard deviation of 2\% of the parameter's nominal value.
    \item   The spectral index of the neutrino flux energy distribution varies introducing a tilt as, $E^{-\gamma +E_{\text{slope}}}$, without constraints. 
   \item The absolute energy scale of the detector, depending on the uncertainty on water optical properties and on the knowledge of PMT efficiencies, as discussed in ref. \cite{NMOpaper}, varies with a standard deviation of 5\% around its nominal value.
   \item $\Delta m^2_{31}$ and $\theta_{23}$ vary without constraints.
   \item $\theta_{13}$ varies with a standard deviation of $0.13^{\circ}$.
\end{enumerate}

The uncertainties used in this work are justified in ref. \cite{NMOpaper}.

\begin{table}[H]
\centering
    \begin{tabular}{|l|c|c|}
      \hline
      \textbf{Parameter} & \textbf{NO} & \textbf{IO}  \\
      \hline
      $\theta_{12}$ & $33.82^{\circ}$ & $33.82^{\circ}$\\
      \hline
      $\theta_{13}$  & $8.60^{\circ}$ & $8.64^{\circ}$ \\
     
      \hline
      $\theta_{23}$ & $48.6^{\circ}$ & $48.8^{\circ}$\\
      \hline
      $\Delta m_{21}^2$ & $7.40 \times 10^{-5}~\mathrm{eV^2} $ & $7.40 \times 10^{-5}~\mathrm{eV^2}$\\ 
      \hline
      $\Delta m_{31}^2$ & $2.528 \times 10^{-3}~\mathrm{eV^2} $ & $-2.436 \times 10^{-3}~\mathrm{eV^2} $ \\ 
      \hline
      $\delta_{\text{CP}}$ & $221^{\circ}$ & $282^{\circ}$\\
      \hline
     
    \end{tabular}
    \caption{List of the three-flavor oscillation parameters from NuFit 4.1 \cite{Esteban_2019} (July 2019) for NO and IO including Super-Kamiokande data. These values are used to generate the observed Asimov datasets ($N^{\text{dat}}_{ij}$). }
    \label{Table2-Nufit}
    \end{table}
   \clearpage
    
    \begin{table}[H]
    \centering
    \begin{tabular}{|c|c|c|}
 \hline
    \textbf{Systematic} & \textbf{Expectation value, $\langle\epsilon_k\rangle$} & \textbf{Standard deviation, $\sigma_k$}\\ \hline
       Track normalisation  & 1 & No prior \\ \hline
        Intermediate normalisation & 1 & No prior\\ \hline
        Shower normalisation  & 1 & No prior \\ 
       \hline
       NC normalisation  & 1 & No prior \\ \hline
       $\nu_{\mu} /\bar{\nu}_{\mu}$ skew ($r_{\mu\bar{\mu}}$) & 0 & 0.05 \\ \hline
       $\nu_{e} /\bar{\nu}_{e}$ skew ($r_{e\bar{e}}$)& 0 & 0.07 \\ \hline
       $\nu_{\mu} / \nu_e$ skew ($r_{\mu e}$)& 0& 0.02 \\ \hline
       $\nu_{\text{up}}/\nu_{\text{hor}}$ skew ($\epsilon_{\theta}$)& 0 & 0.02 \\ \hline
       Energy scale & 1 & 0.05\\ \hline
       Spectral index ($E_{\text{slope}})$ & 0 & No prior \\ \hline

        $ \Delta m^2_{31} (\text{NO/IO}) [10^{-3}~\text{eV}^2]$ & $2.528/-2.436 $ & No prior \\ \hline
 
      $\theta_{23} (\text{NO/IO}) [^{\circ}]$ & $48.6/48.8$ & No prior \\ \hline
             $\theta_{13} (\text{NO/IO}) [^{\circ}]$  & $8.60/8.64$ & $0.13$ \\ \hline

    \end{tabular}
    \caption{List of systematic parameters used in the fitted model with their corresponding priors. Their expectation values are used to generate the observed Asimov dataset ($N^{\text{dat}}_{ij}$) and they are fitted when computing the number of expected events in the model ($N^{\text{mod}}_{ij}$).  }
    \label{Table3-Nuisance}
\end{table}

\section{Results}
\label{sec:results}
In this section, the potential of KM3NeT/ORCA to constrain the invisible neutrino decay is reported and compared with the limits and sensitivities published by current and future experiments. The possibility to exclude the stable neutrino hypothesis assuming the maximum $\alpha_3$ allowed by current limits is discussed. Afterwards, a discussion on the effect of the actual value of $\theta_{23}$ on the sensitivity and the discovery potential of neutrino decay is included. Finally, the influence of the decay on the measurement of the standard oscillation parameters is presented.

\subsection{Invisible neutrino decay sensitivity}
\begin{figure}[H]
    \centering
    \subfloat{
   \label{Fig5_Sensitivity_sum}
     \includegraphics[height=6cm]{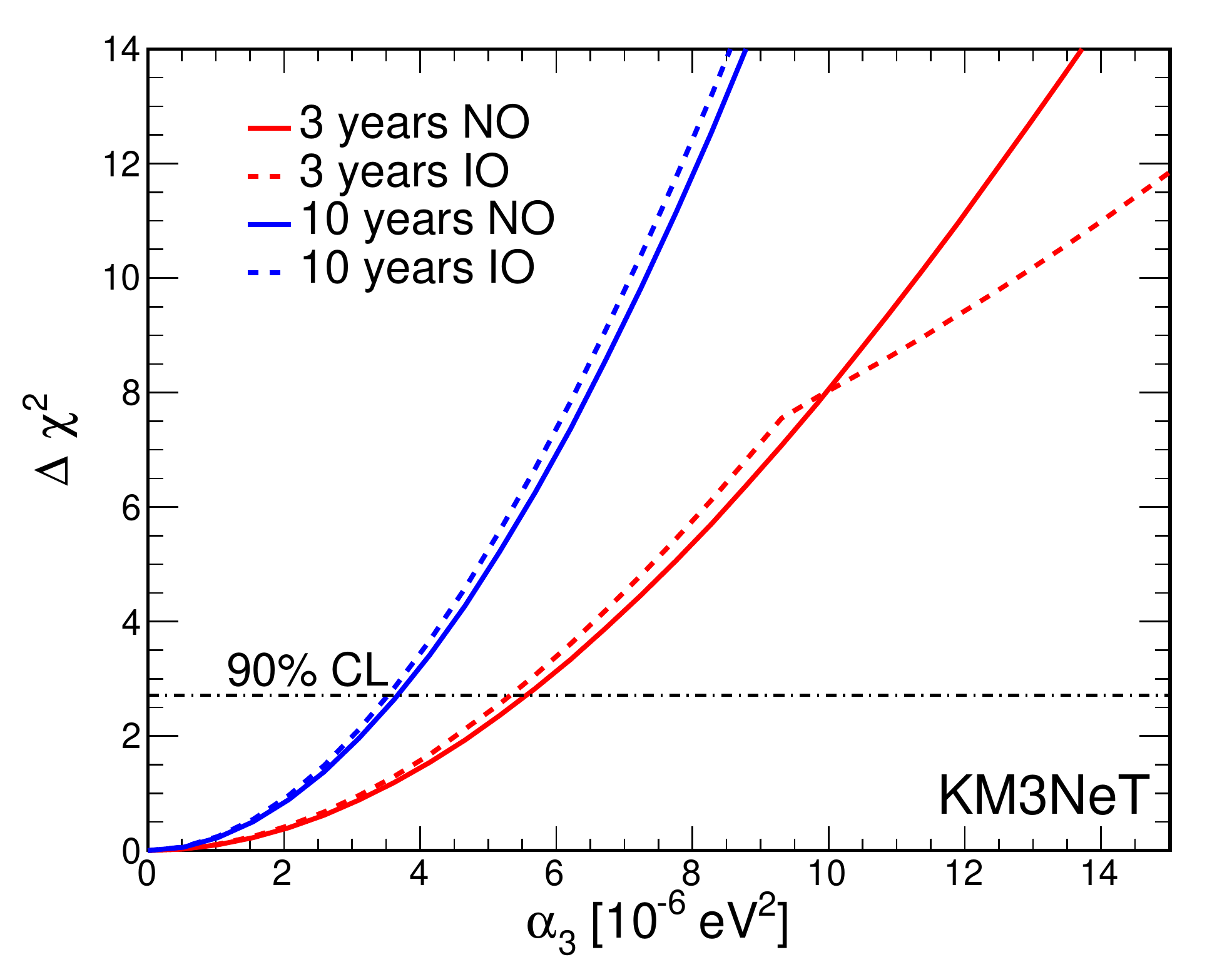}}
      \subfloat{
   \label{Fig5_Sensitivities_split}
     \includegraphics[height=6cm]{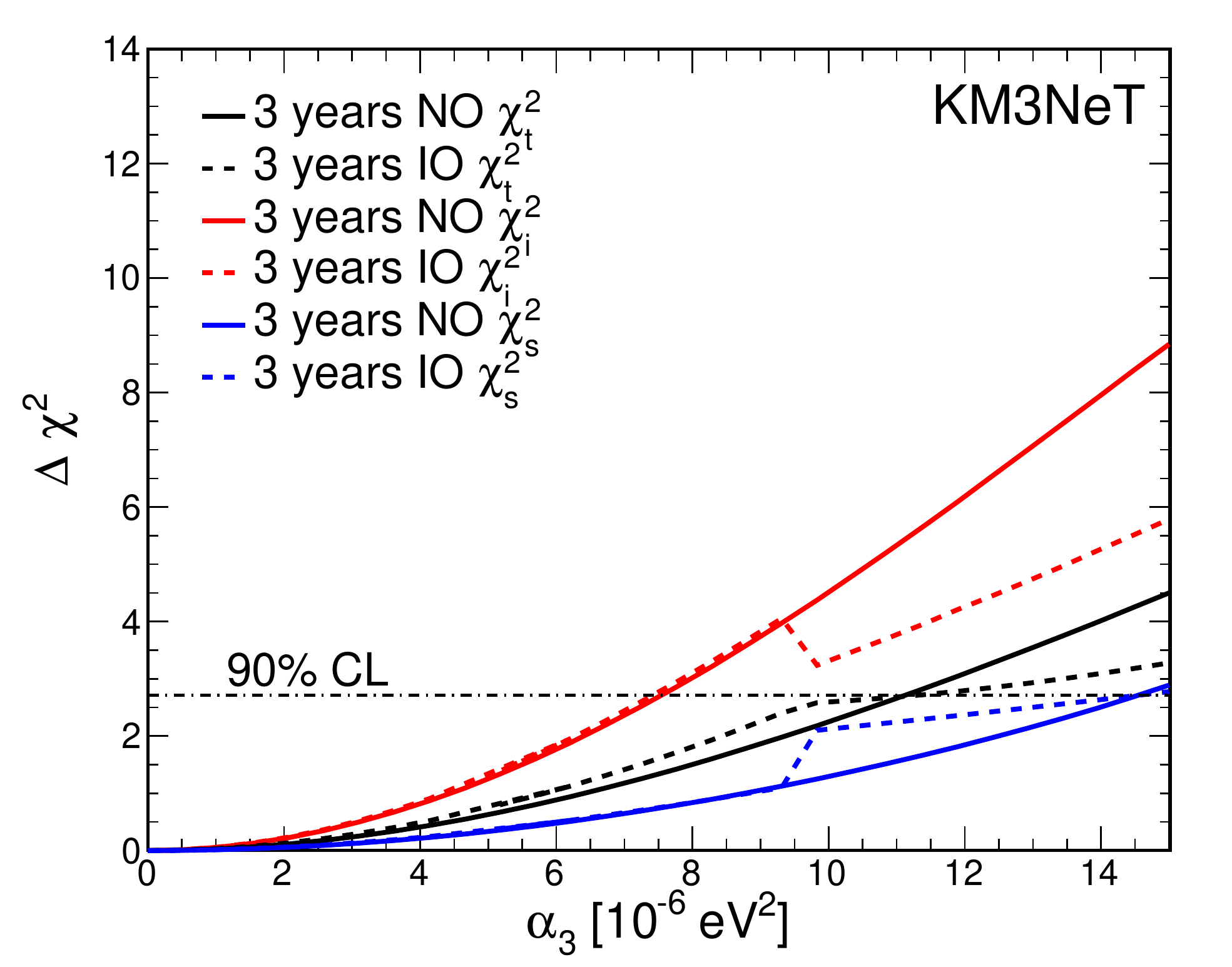}}
    \caption{Left: Sensitivity to invisible neutrino decay for 3 (red) and 10 (blue) years of data taking in KM3NeT/ORCA assuming NO (solid lines) and IO (dashed lines) as a function of $\alpha_3$ in the model. Right: Sensitivity to invisible neutrino decay assuming NO (solid lines) and IO (dashed lines) for 3 years of data taking as a function of $\alpha_3$ in the model per topology: $\chi^2_t$ for track events (black), $\chi^2_i$ for intermediate events (red), $\chi^2_s$ for shower events (blue).
}
    \label{Fig5_Sensitivity}
\end{figure}

Figure \ref{Fig5_Sensitivity} shows the KM3NeT/ORCA sensitivity to the invisible neutrino decay for 3 and 10 years of data taking assuming true normal or inverted ordering. In both cases the sensitivities for true IO present a change in the slope of the $\chi^2$ around $\alpha_3=10^{-5}~\mathrm{eV^2}$ resulting in a degradation of the sensitivity. The sensitivity for IO gets worsened due to correlations with $\theta_{23}$ previously mentioned in section \ref{sec:oscframe}. Higher values of $\alpha_3$ in the tested model with respect to the true $\theta_{23}$ used to simulate the Asimov dataset are compensated by flipping $\theta_{23}$ to the lower octant in the model. This is differently observed when true NO is assumed because the change in the slope is softer and subtle. In figure \ref{Fig5_Sensitivity} right, the $\chi^2$ contribution from the track-like, intermediate and shower-like samples for 3 years of data taking are shown for better understanding of this behaviour. The $\chi^2$ contribution from the intermediate sample is reduced, while the contribution from the shower sample is increased when the octant is flipped. This will be further developed in section \ref{sec:th23}. The 90\% confidence level (CL) upper limits obtained for 3 and 10 years of data taking are shown in table \ref{Table4-results}.

\begin{table}[H]
    \centering
    \begin{tabular}{|c|c|c|}
    \hline
    \centering
       Data taking [yr]    & $\alpha_3$ NO [$10^{-6}~\mathrm{eV^2}$] & $\alpha_3$ IO [$10^{-6}~\mathrm{eV^2}$] \\  \hline 
  
   3 & 5.7 & 5.3 \\\hline
   10 & 3.7 & 3.5 \\ \hline
    \end{tabular}
    \caption{90\% CL Upper limits for the decay constant, $\alpha_3$, for 3 and 10 years of data taking for both neutrino mass ordering hypotheses.}
   \label{Table4-results}
\end{table}

The corresponding lower limits at $90\%$ CL in terms of the lifetime parameter, $\tau_3/m_3$, are shown in table \ref{Table5_AllExp}, where the bounds on $\tau_3/m_3$ of current (blue) and future experiments, are also collected. After 10 years of data taking, KM3NeT/ORCA will improve the current bounds on the invisible neutrino decay by two orders of magnitude with respect to existing long-baseline results, and it will be at least, as competitive as future experiments, such as DUNE, JUNO, MOMENT, ESSnuSB, and INO. Note that
the limit coming from K2K, MINOS, SK I+II combination \cite{SKK2KMINOS} was derived under the two-neutrino approximation and without matter effects,  therefore, a direct comparison is not possible without an updated analysis. 

A previous estimation of the KM3NeT/ORCA sensitivity to invisible neutrino decay was published in ref. \cite{de_Salas_2019} showing that a lower limit at  $90\%$ CL of $\tau_3/m_3>250~\text{ps/eV}$ could be reached after 10 years of data taking. The analysis presented in this paper incorporates improvements in the neutrino detection efficiency, reconstruction performance and more refined analysis methods. In particular, the correlations between the reconstructed variables, as well as, the overestimations of the sensitivity due to MC statistics are taken into account and a more complete set of systematic effects is considered. These improvements result in a more reliable sensitivity calculation.

\begin{table}[H]
    \centering
    \begin{tabular}{|l|c|c|c|}
    \hline
    Experiment   &   UL ($90\%$ CL) [$\mathrm{10^{-6} eV^2}$] & LL ($90\%$ CL) [ps/eV]   & Reference\\
    \hline
      \textbf{KM3NeT/ORCA (3 yr)} & 5.7 & \textbf{$120 $}  & This work\\ \hline
      \textbf{KM3NeT/ORCA (10 yr)} & 3.7 & \textbf{$180 $}  & This work\\ \hline
 
    \textcolor{blue}{T2K, NOvA} &\textcolor{blue}{290} &\textcolor{blue}{$2.3 $} &   \cite{NovaT2K} \\ \hline
    \textcolor{blue}{T2K, MINOS} &\textcolor{blue}{240} &\textcolor{blue}{$2.8 $} &   \cite{MinosT2K} \\ \hline
   \textcolor{blue}{K2K, MINOS, SK I+II} &\textcolor{blue}{2.3} & \textcolor{blue}{$290 $} &  \cite{SKK2KMINOS}\\ \hline
        MOMENT (10 yr) & 24&$28 $  &\cite{MOMENT_Tang_2019} \\ \hline
  
     ESSnuSB (5$\nu$+5$\bar{\nu}$) yr &$16-13$ &$42-50 $ & \cite{ESSnuSB_choubey2020exploring} \\  \hline
         DUNE (5$\nu$+5$\bar{\nu}$) yr &13 &$51$ &  \cite{DUNEupdated} \\ \hline
    JUNO (5 yr) & 7&$93 $ &  \cite{abrahao2015constraint} \\  \hline
    INO-ICAL (10 yr) & 4.4&$151$ & \cite{Choubey_2018_INO} \\ \hline
    \end{tabular}
  \caption{ Upper limits (UL) and their corresponding conversion into lower limits (LL) at $90\%$ CL for the decay constant, $\alpha_3$, and its inverse, $1/\alpha_3=\tau_3/m_3$,  for current (blue) and future experiments.}
  \label{Table5_AllExp}
\end{table}

\subsection{Invisible neutrino decay discovery potential}

\begin{figure}[H]
    \centering
   
     \includegraphics[height=6cm]{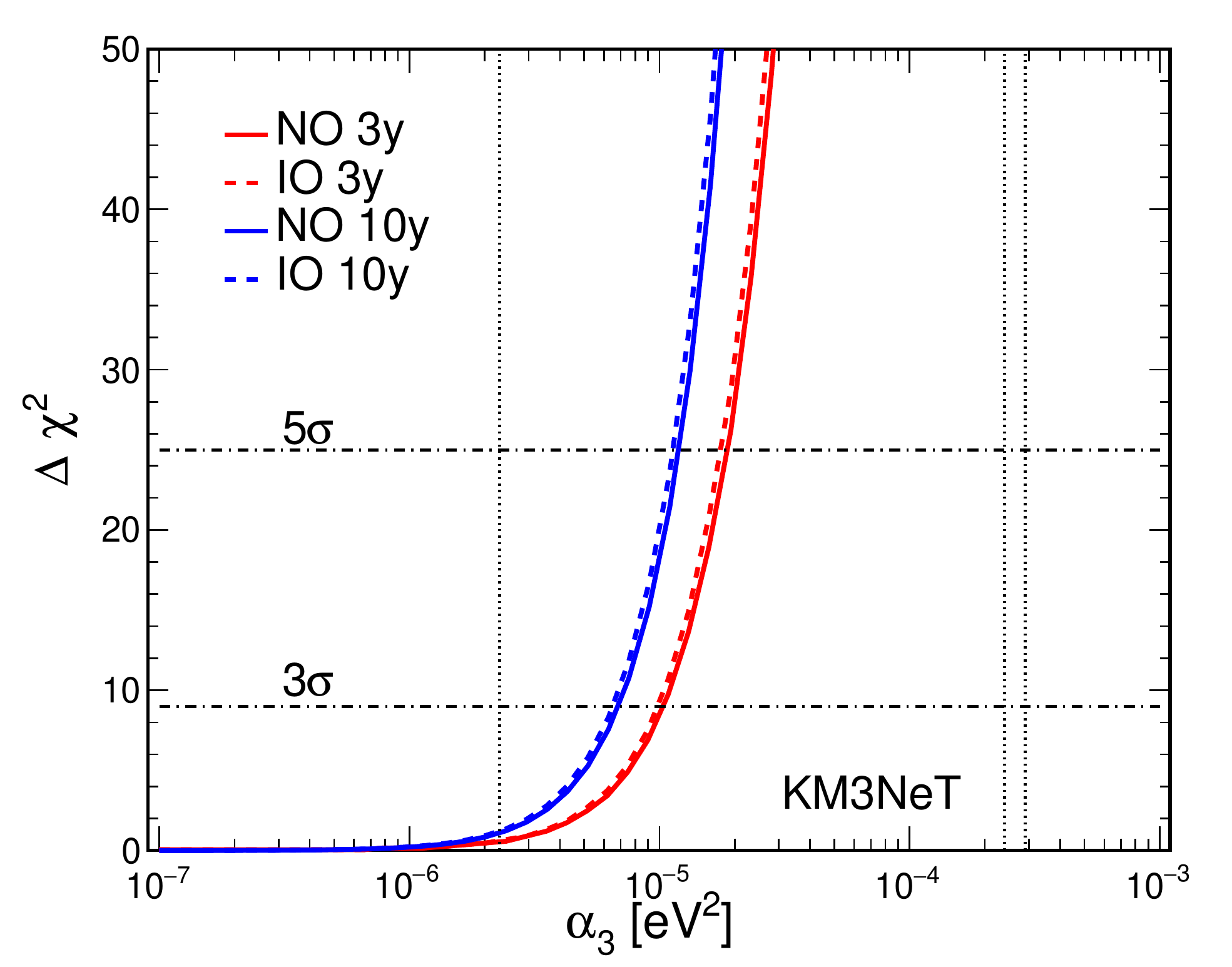}
    \caption{Discovery potential for 3 (red) and 10 (blue) years of data taking in KM3NeT/ORCA assuming NO (solid lines) and IO (dashed lines) as a function of the true $\alpha_3$ in the Asimov dataset. The vertical lines are the 90$\%$ CL limit $\alpha_3$ values of current experiments reported in table \ref{Table5_AllExp}. From left to right: SK+K2K+MINOS, T2K+MINOS and T2K+NO$\nu$A.}
    \label{Fig6_Discovery}
\end{figure}

In this section, the capability of KM3NeT/ORCA to discover invisible neutrino decay and exclude the stable neutrino hypothesis is explored. The observed Asimov dataset, $N_{ij}^{\text{dat}}$, is simulated assuming as true
value of $\alpha_3$ a wide range of values while the model is fitted assuming standard oscillations. In the case the true invisible decay parameter value was $\alpha_3=2.4 \times 10^{-4}~\mathrm{eV^2}$ (T2K+MINOS limit) or $\alpha_3=2.9 \times 10^{-4}~\mathrm{eV^2}$ (T2K+NO$\nu$A limit), KM3NeT/ORCA could rule out the hypothesis of stable neutrinos with a significance above $5\sigma$ as shown in figure \ref{Fig6_Discovery}. In the case the true invisible decay parameter value was $\alpha_3=2.3 \times 10^{-6}~\mathrm{eV^2}$ (SK+K2K+MINOS limit) the significance is below 3$\sigma$. In table \ref{Table6-DiscoveryI} the values KM3NeT/ORCA could exclude with $3\sigma$ or $5\sigma$ are shown for both mass ordering hypotheses and 3 and 10 years of data taking.
\clearpage

\begin{table}[H]
    \centering
    \begin{tabular}{|c|c|c|}
    \hline
    \centering
       Significance    & 3y NO/IO [$10^{-6}~\mathrm{eV^2}$] & 10y NO/IO [$10^{-6}~\mathrm{eV^2}$] \\  \hline 
  
   3$\sigma$ & 10.5/10.3 & 6.8/6.5 \\\hline
   5$\sigma$ & 18.6/17.8 & 12.0/11.4 \\ \hline
    
    \end{tabular}
    \caption{Discovery potential for 3 and 10 years of data taking for both NMO hypotheses. }
   \label{Table6-DiscoveryI}
\end{table}

\subsection{The role of $\theta_{23}$ in the invisible neutrino decay}

\label{sec:th23}

\begin{figure}[H]
 \centering
  \subfloat{
   \label{Fig7-StatTheta23-1}
     \includegraphics[height=6cm]{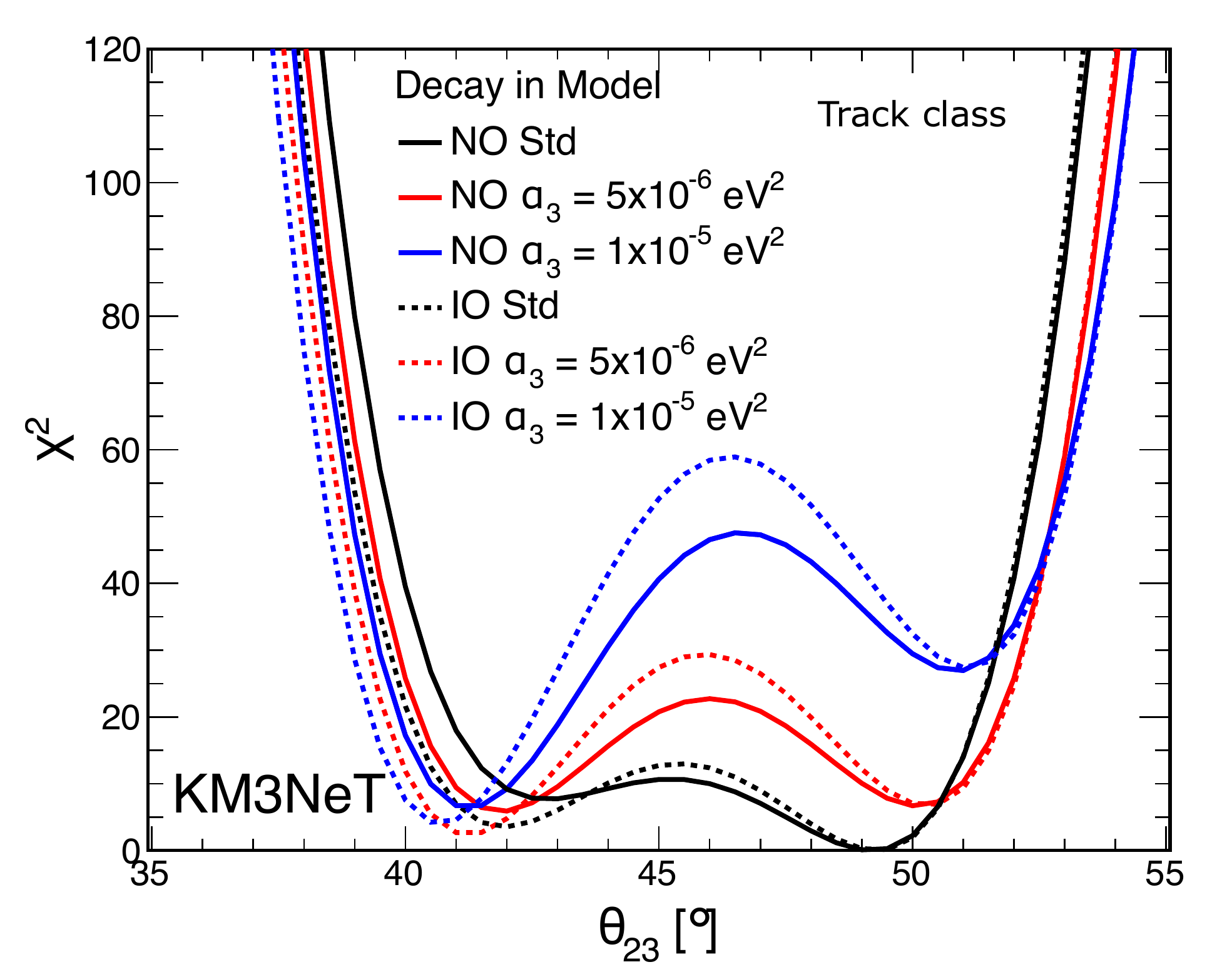}}
      \subfloat{
   \label{Fig7-StatTheta23-2}
     \includegraphics[height=6cm]{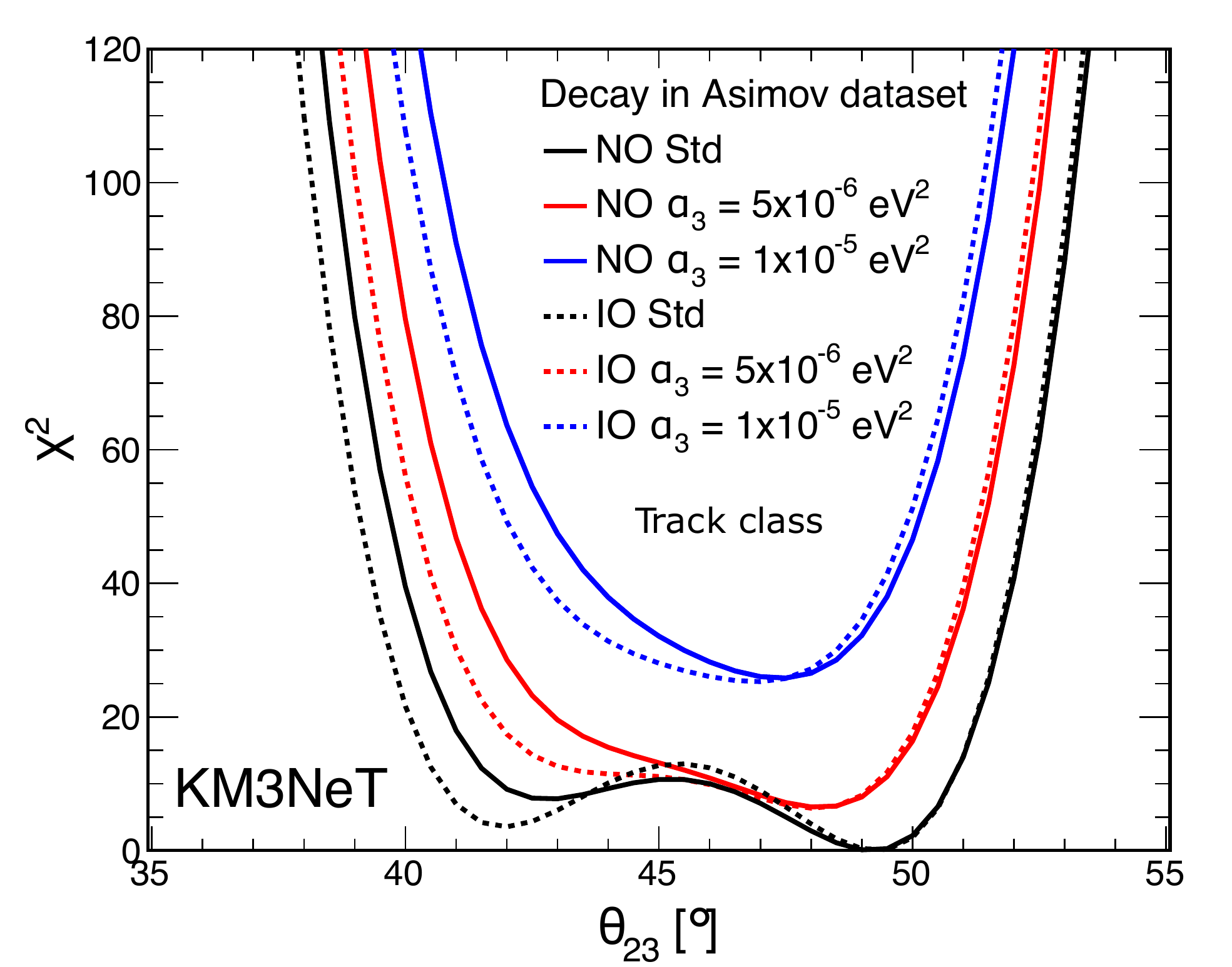}} \\
  \subfloat{
   \label{Fig7-StatTheta23-3}
      \includegraphics[height=6cm]{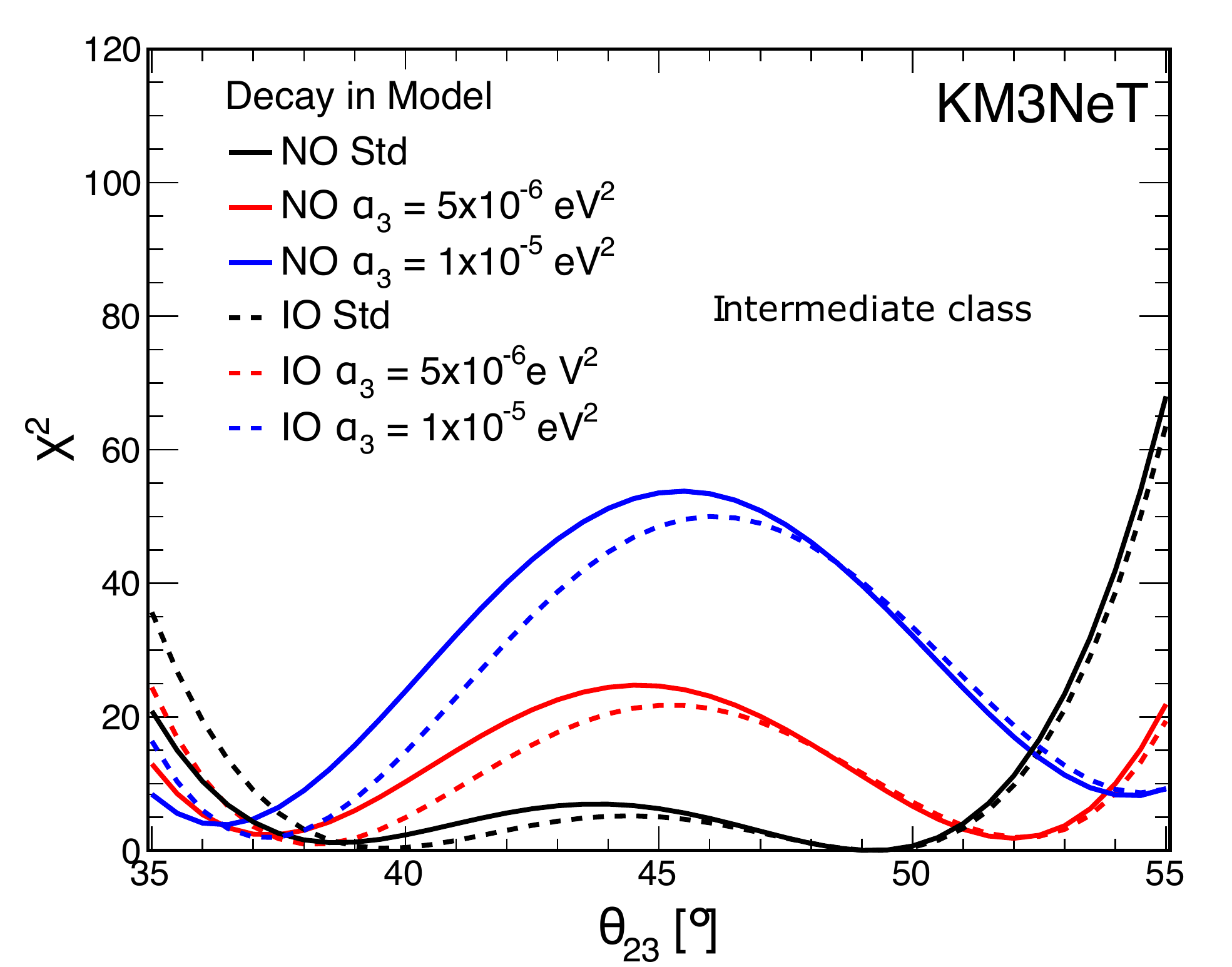}}  \subfloat{
   \label{Fig7-StatTheta23-4}
      \includegraphics[height=6cm]{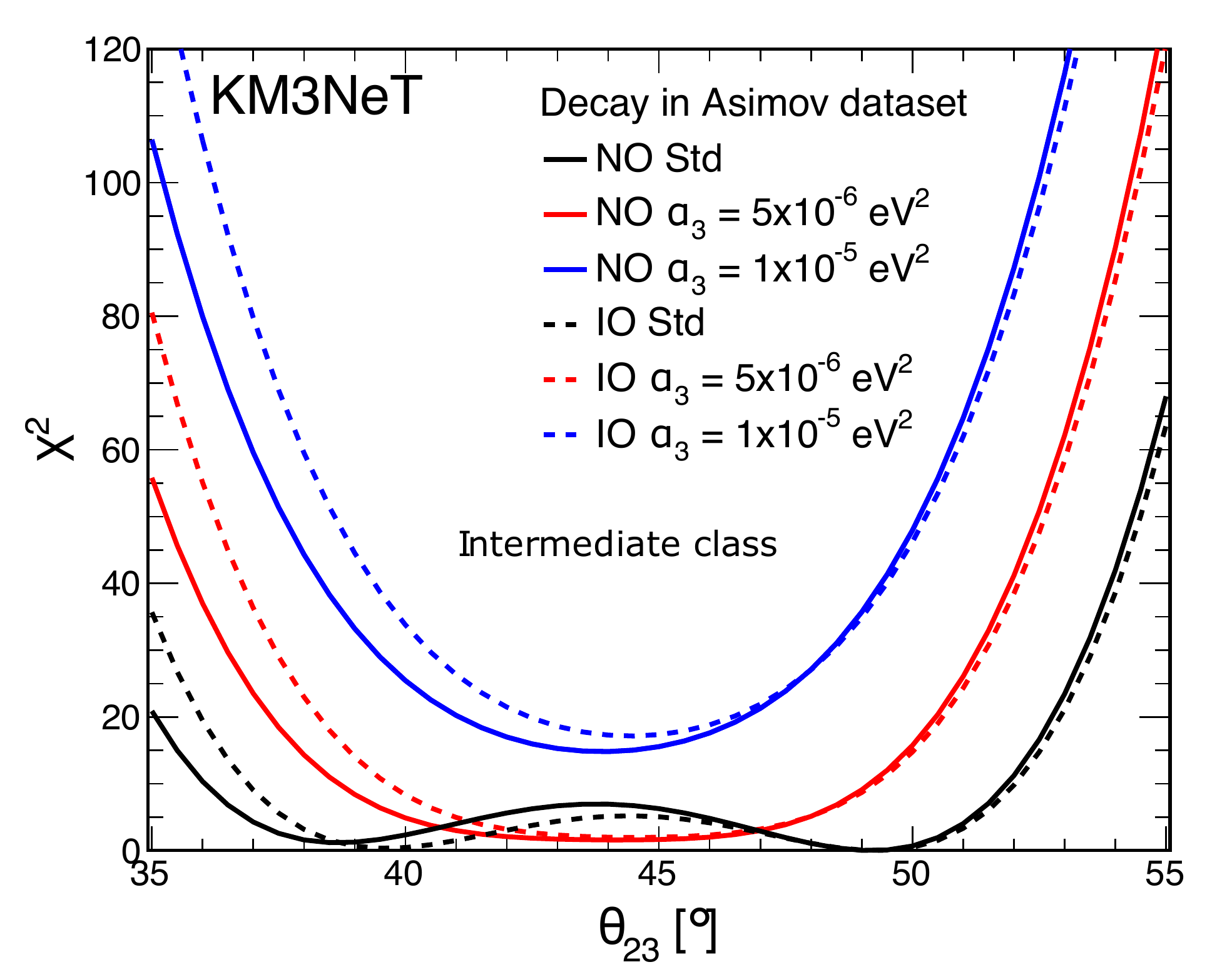}} \\
       \subfloat{
   \label{Fig7-StatTheta23-5}
      \includegraphics[height=6cm]{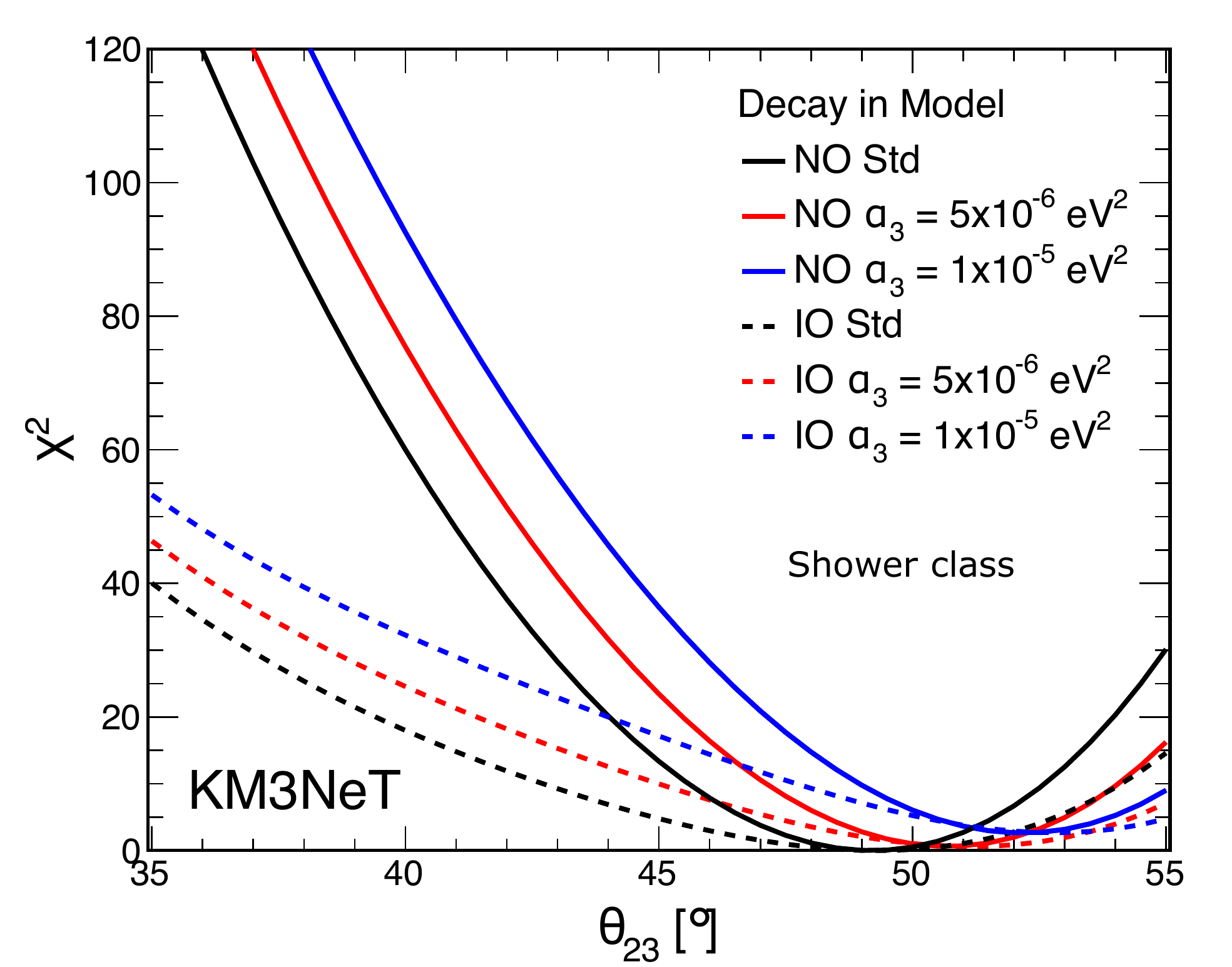}} 
      \subfloat{
   \label{Fig7-StatTheta23-6}
      \includegraphics[height=6cm]{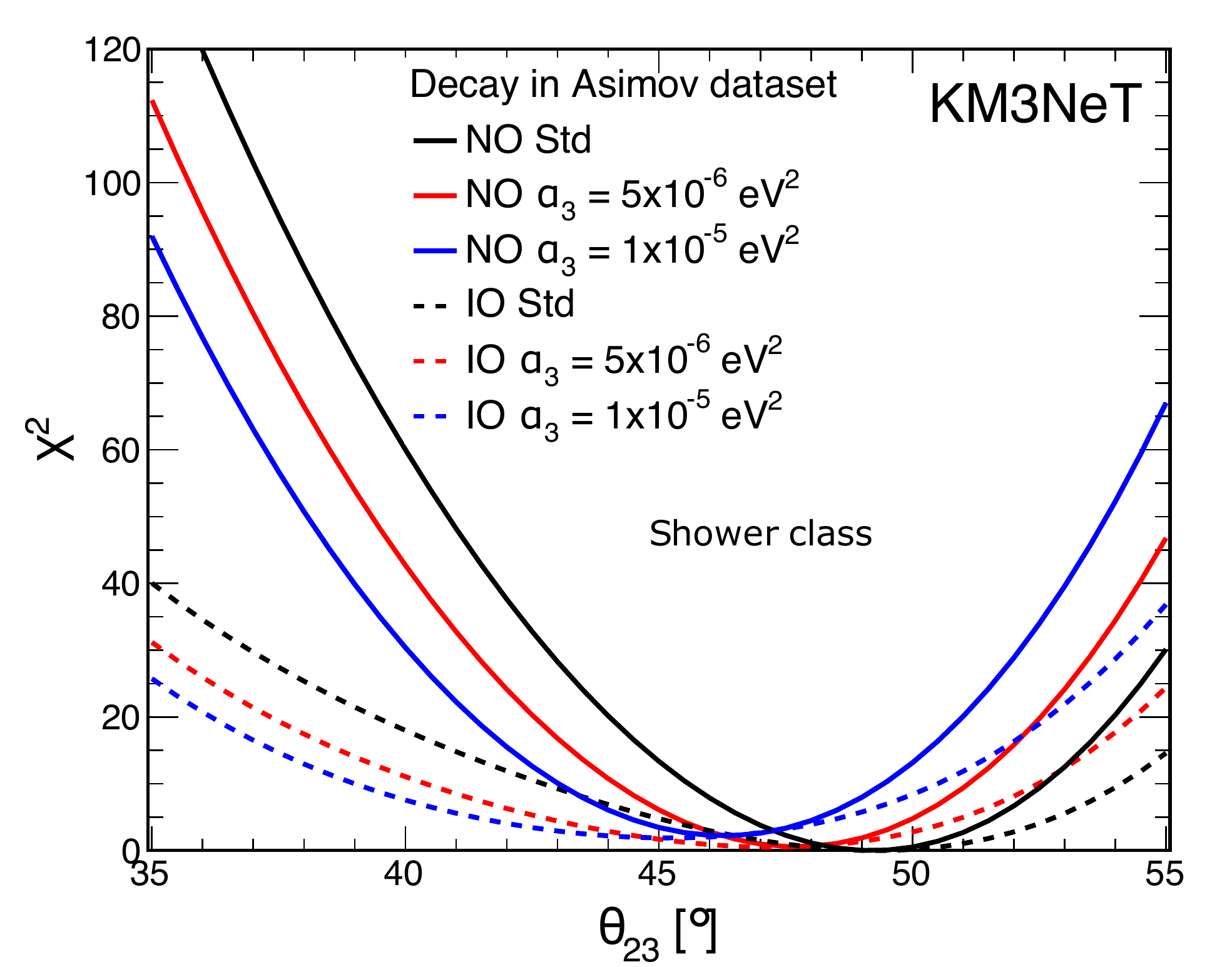}} 
 \caption{Statistical $\chi^2$ as a function of the scanned value of $\theta_{23}$ in the model for different values of $\alpha_3$ fixed in the model (left) or in the Asimov dataset (right) for NO (solid) and IO (dashed) separated by class topology contributions: track-like (top), intermediates (middle) and shower-like (bottom). Three cases are considered, $\alpha_3=0$ (black), $\alpha_{3}=5\times 10^{-6}~\mathrm{eV^2}$ (red) and $\alpha_{3}=10^{-5}~\mathrm{eV^2}$ (blue).}
  \label{Fig7-StatTheta23}
\end{figure}

\begin{figure}[H]
 \centering
   \subfloat{
   \label{Fig8-StatSum-1}
     \includegraphics[height=6cm]{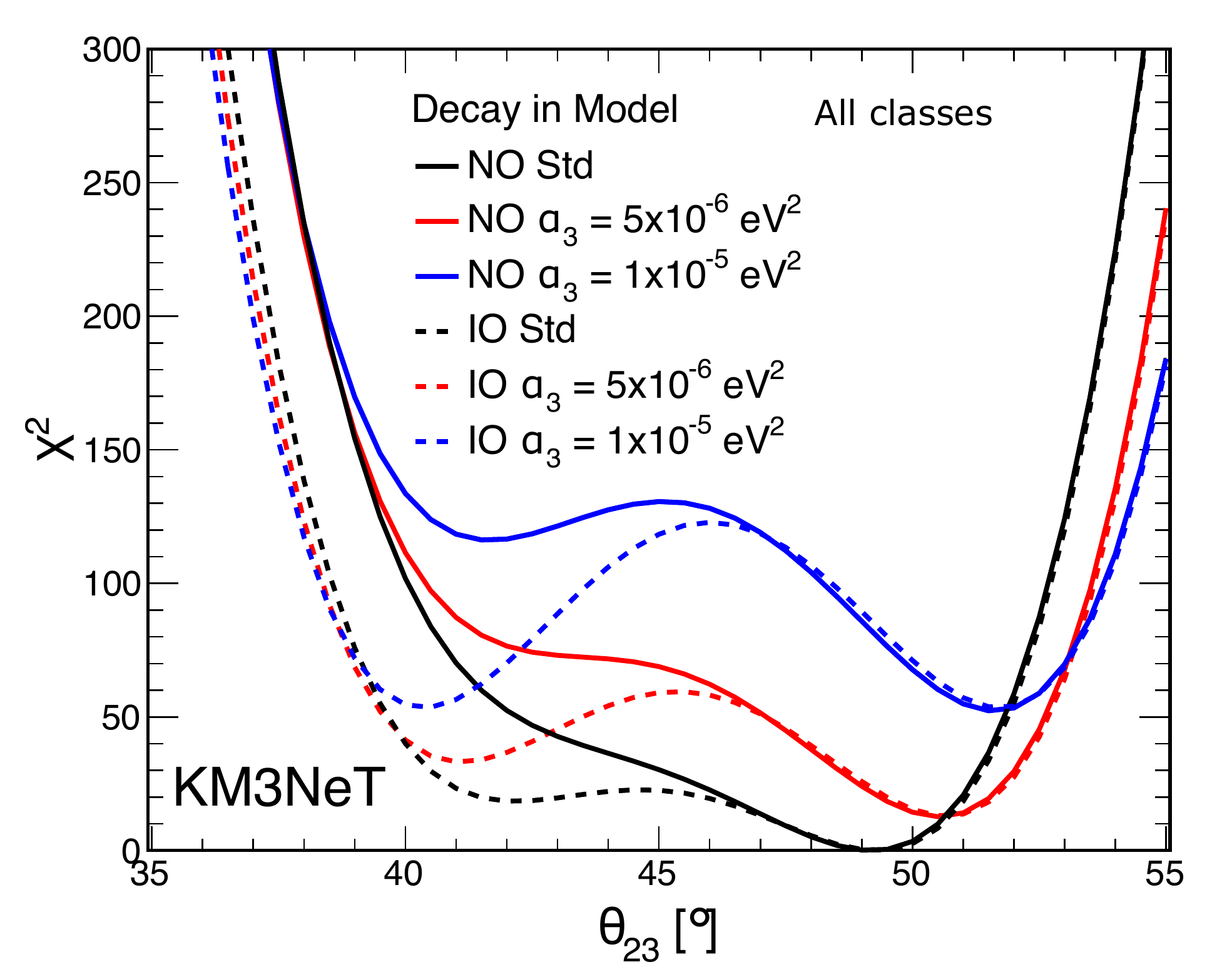}}
      \subfloat{
   \label{Fig8-StatSum-2}
     \includegraphics[height=6cm]{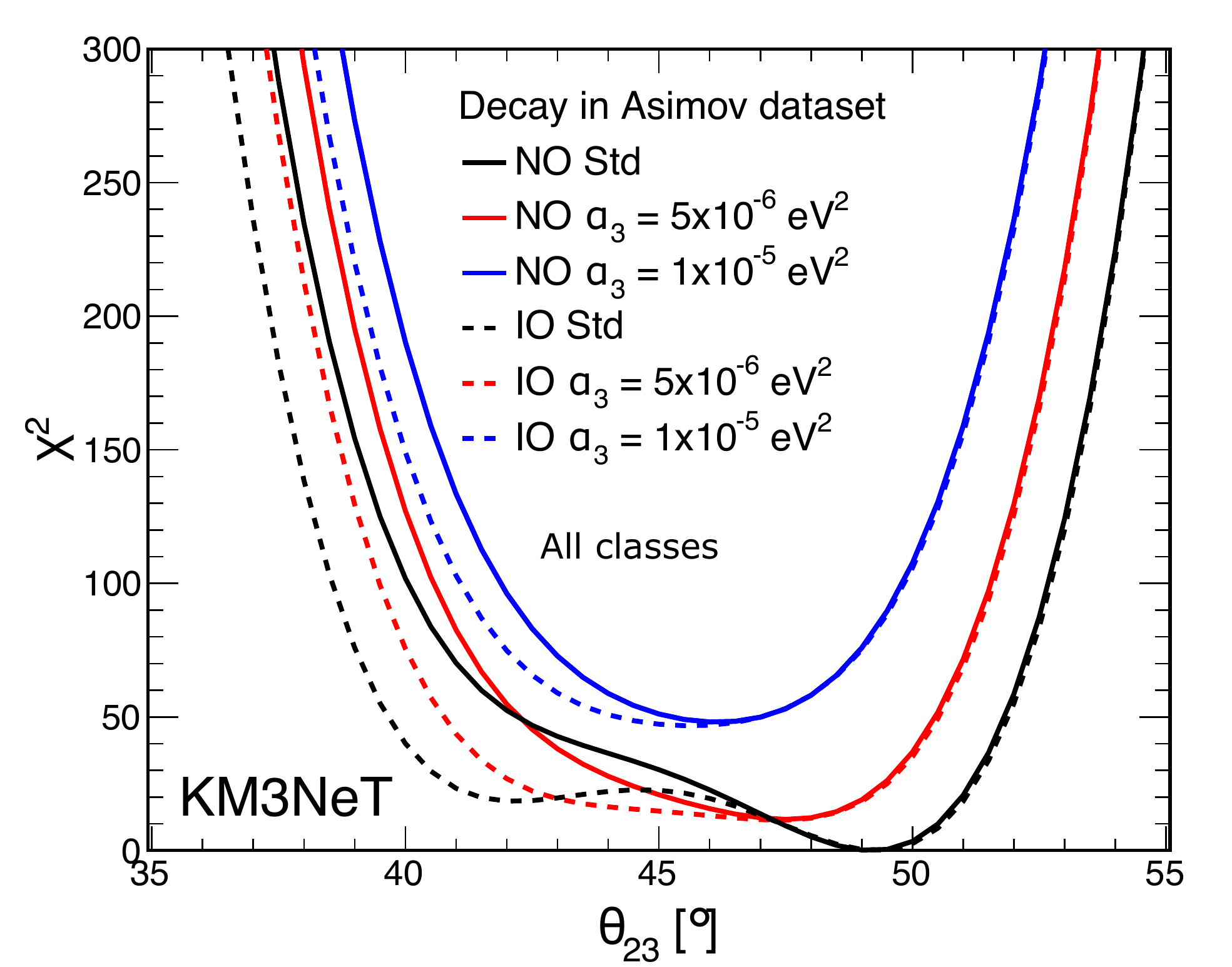}}
    
 \caption{Statistical $\chi^2$ as a function of the scanned value of $\theta_{23}$ in the model for different values of $\alpha_3$ fixed in the model (left) and in the Asimov dataset (right) for NO (solid) and IO (dashed) with all contributions summed up together. Three cases are considered, $\alpha_3=0$ (black), $\alpha_{3}=5\times 10^{-6}~\mathrm{eV^2}$ (red) and $\alpha_{3}=1\times 10^{-5}~\mathrm{eV^2}$ (blue).}
  \label{Fig8-StatSum}
\end{figure}

The worsening of the sensitivity (figure \ref{Fig5_Sensitivity}) in the case of IO is related to the octant flip of $\theta_{23}$. In order to understand this behaviour, figure \ref{Fig7-StatTheta23} shows the dependence of the statistical sensitivity (left panels) and discovery (right panels) as a function of the tested $\theta_{23}$ in the model for the three topology classes and for three values of the decay constant. In the sensitivity plots, decay is assumed in the model and the Asimov dataset is generated with standard oscillations for a true value of $\theta_{23}= 48.6^{\circ}$.  The main contribution to the track-like class comes from $P_{\mu\mu}$ which in the absence of decay has a non-negligible preference to the higher $\theta_{23}$ octant. However, in the presence of decay, the preference starts to shift to the lower octant. There is a clear preference for the lower octant for the track-like class (figure \ref{Fig7-StatTheta23} top left). The small differences between NO and IO come from the contribution of the matter-enhanced transition probabilities $P_{e\mu}$, which are more suppressed for IO. Although the shower-like class  is dominated mainly by $P_{ee}$, which is $\theta_{23}$ blind, it is nevertheless able to measure the octant trough $P_{\mu e}$. Therefore, a shift to the lower octant is prevented. On the other hand, higher values of $\theta_{23}$ are favoured to compensate the loss due to decay (figure \ref{Fig7-StatTheta23} bottom left). In this case, however, the difference between NO and IO is larger, because the sensitivity in the determination of the $\theta_{23}$ octant comes from matter-resonance effects in the transition probabilities: in particular, from neutrinos in the case of NO and from antineutrinos in the case of IO. Since in KM3NeT/ORCA more neutrino events than antineutrino events  are expected, the $\chi^2$  increases more steeply for NO than for IO. This is the reason why, when adding up all contributions of the left plots in figure \ref{Fig8-StatSum}, as long as $\alpha_3$ increases, the preferred octant will be the lower one. This behaviour is more pronounced in the case of true IO as  shown in figure \ref{Fig8-StatSum} left.

For the discovery potential (right panels of figure \ref{Fig7-StatTheta23}),  the Asimov dataset is simulated with a non-zero value of $\alpha_3$ and $\theta_{23} =48.6^{\circ}$ while standard oscillations are assumed in the model.  Since there is no decay in the model, the effects of changing the value of $\theta_{23}$ affect mainly transition channels, which are proportional to $\sin^2(\theta_{23})$, so to reduce the number of events induced by the decay in data, the model favors lower values of  $\theta_{23}$ for all topological classes. In figure \ref{Fig8-StatSum} right, the three contributions are summed up together. 
\clearpage

\begin{figure}
 \centering
      \includegraphics[height=7cm]{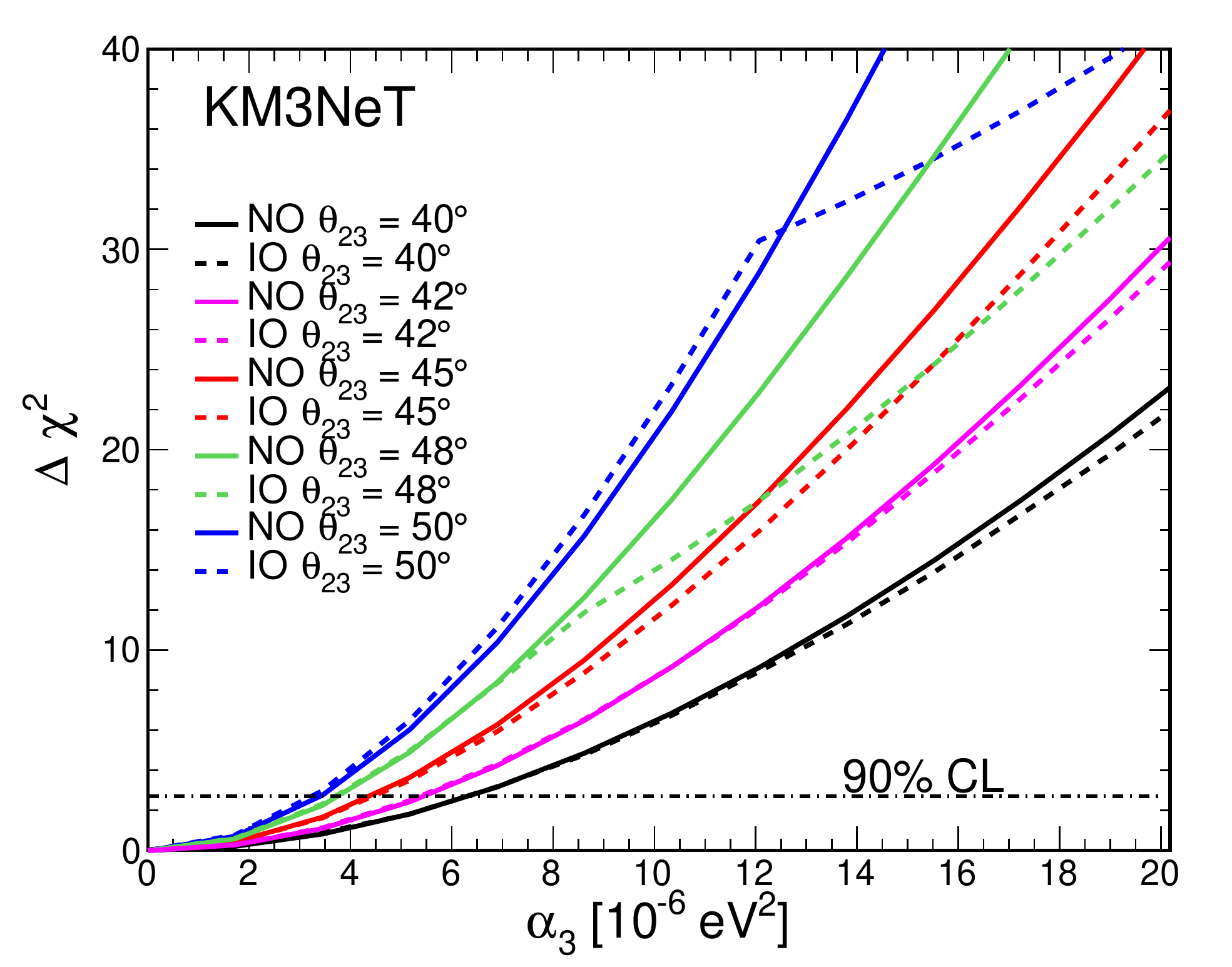}
 \caption{Sensitivity to neutrino decay for 10 years of data taking in KM3NeT/ORCA assuming NO (solid) and IO (dashed) as a function of $\alpha_3$ for different values of the true $\theta_{23}$.}
 \label{Fig9-SensitivityThetaTrue}
\end{figure}

\begin{figure}[]
 \centering
    \includegraphics[height=7cm]{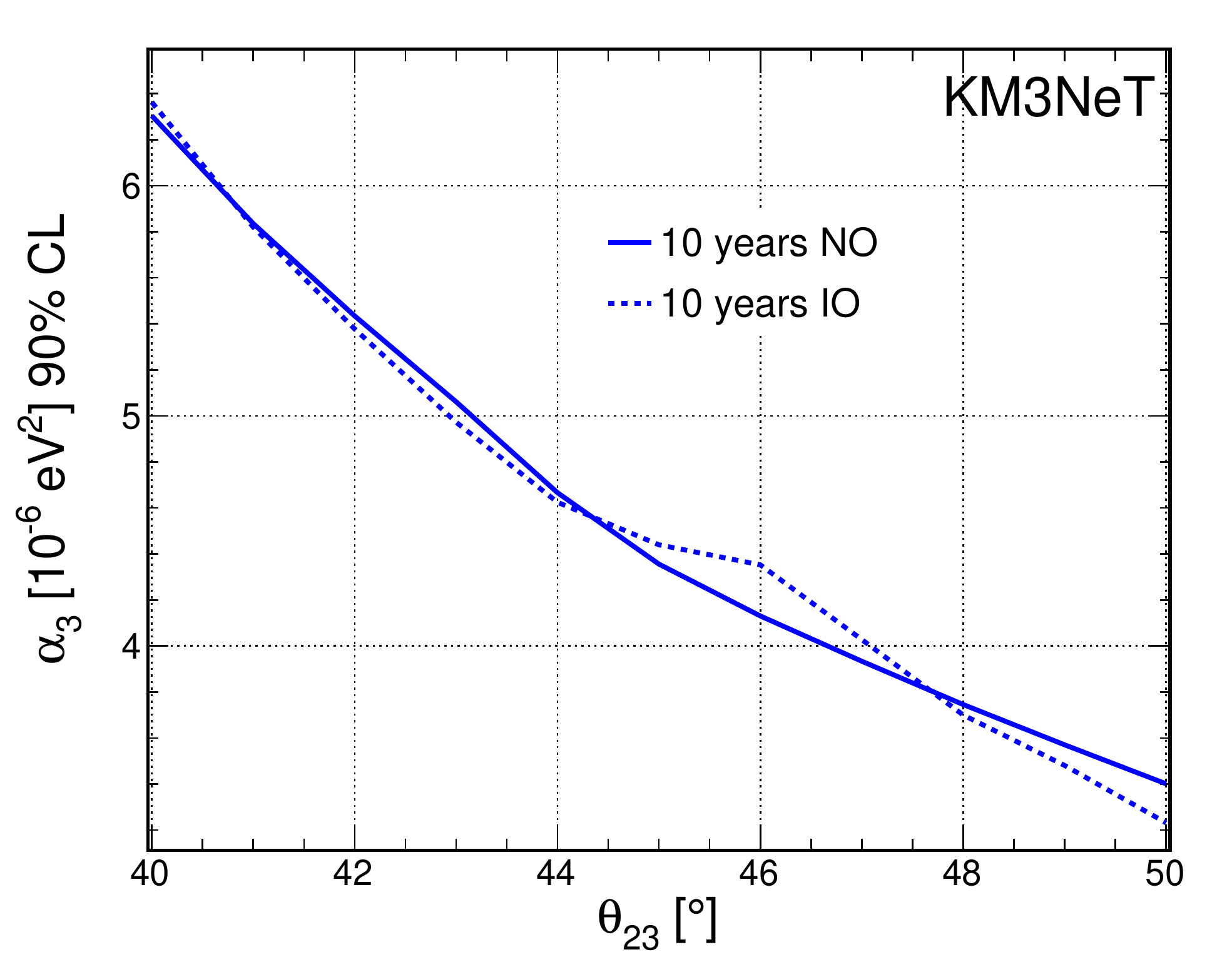} 

        \caption{$90\%$ CL upper limits for $\alpha_3$ in KM3NeT/ORCA after 10 years of data taking assuming NO (solid) and IO (dashed) as a function of the true value of $\theta_{23}$. }
                \label{Fig10-Sensitivity90}
 \end{figure}

The total expected sensitivity including systematic effects is also affected by the true $\theta_{23}$ value used to generate the Asimov dataset. Figure \ref{Fig9-SensitivityThetaTrue} shows the KM3NeT/ORCA sensitivity to invisible neutrino decay for different true values of $\theta_{23}$. The lower the true value of $\theta_{23}$, the lower the sensitivity for  $\alpha_3$. Since sensitivity to neutrino decay around the matter-resonant energy region depends mainly  on the transition channels, which  are proportional to $\sin^2(\theta_{23})$, a true lower value of $\theta_{23}$ reduces the number of events, and thus, the ability of KM3NeT/ORCA to constrain invisible neutrino decay.
 Figure \ref{Fig10-Sensitivity90} shows the  $90\%$ CL upper limit on $\alpha_3$ as a function of the true $\theta_{23}$. True $\theta_{23}$ values in the upper octant result in degrading the sensitivity for true IO when the octant is flipped in the fit.

The sensitivity on $\alpha_3$ can change by up to a factor of 2 when moving the true $\theta_{23}$ from $40^{\circ}$ to $50^{\circ}$. Differences between NO and IO are small below $45^{\circ}$ but they increase when approaching maximal mixing. Around $46^{\circ}$ the sensitivity gets worse for IO than for NO. This effect appears because around $45^{\circ}$ the octant flip of the profiled $\theta_{23}$ happens for lower values of the decay constant. For 10 years of data taking, lower limits on $1/\alpha_3=\tau_3/m_3$ range between $100$ ps/eV (the worst scenario, $\theta_{23}=40^{\circ}$) and $190$ ps/eV (the best scenario, $\theta_{23}=50^{\circ}$). 

\begin{figure}[H]
 \centering

   \label{CNO}
     \includegraphics[height=5.8cm]{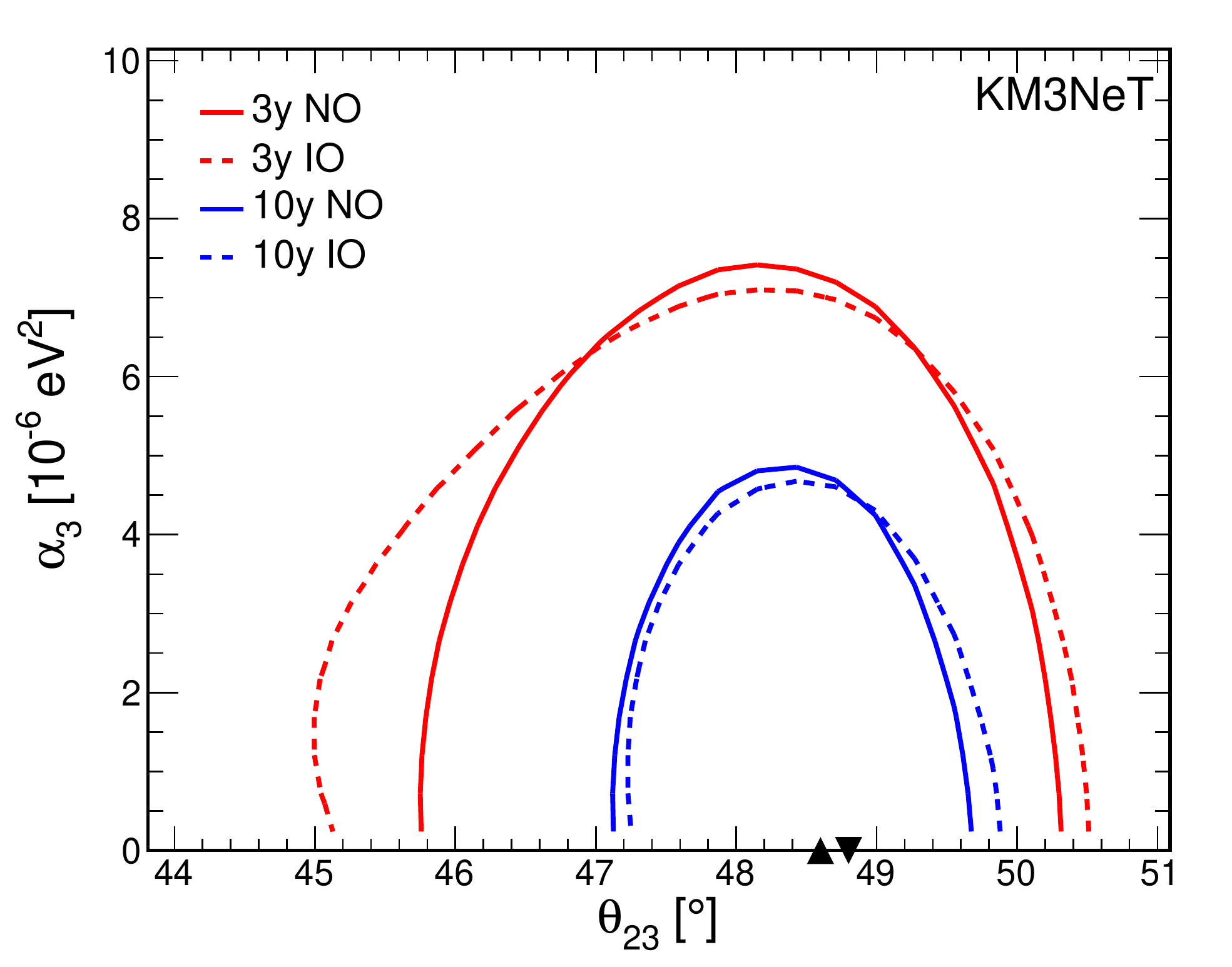}

        \caption{Contours in the $\theta_{23} - \alpha_3$ plane at 90\% CL for both NMO hypotheses for 3 (red) and 10 (blue) years of data taking. The true values used to simulate the Asimov dataset are marked with triangles, pointing up for NO and pointing down for IO. }
\label{Fig11_Contour_TA}
 \end{figure}

The ability of KM3NeT/ORCA to constrain both $\alpha_3$ and $\theta_{23}$ simultaneously can be seen in figure \ref{Fig11_Contour_TA}, where the 90\% CL contour plot in the $\alpha_3$-$\theta_{23}$ plane is shown for 3 and 10 years of data taking. For IO, a marginal difference  appears for lower values of $\theta_{23}$ that are allowed for IO but not for NO at $90\%$ CL for 3 years of data taking. This effect is not present for 10 years of data taking. The potential of KM3NeT/ORCA to constrain $\theta_{23}$ is high enough to avoid degeneracies at this CL \cite{NMOpaper}.
\clearpage

 \subsection{The impact of invisible neutrino decay on the oscillation parameter measurements.}

 \label{sec:Cases}
The presence of  neutrino decay could affect the measurement of the oscillation parameters such as $\theta_{23}$ and $\Delta m^2_{31}$. In order to study this possible impact, expected CL contours in the $\theta_{23}-\Delta m^2_{31}$ plane have been computed in the following cases:

\begin{enumerate}
    \item Both Asimov dataset and tested model are generated with standard oscillations. 
    \item Asimov dataset is generated with standard oscillations and it is analysed assuming a model where the neutrino can decay, leaving $\alpha_3$ as a free parameter.
    \item Asimov dataset is generated  assuming  current limits in neutrino decay ($\alpha_3=1.22\times 10^{-5}~\mathrm{eV}^2$, current best limit at
$3\sigma$ from the SK+K2K+MINOS combined analysis\cite{SKK2KMINOS}) and $\alpha_3$ is left as a free parameter in the model.
    \item Asimov dataset is generated assuming  current limits in neutrino decay ($\alpha_3=1.22\times 10^{-5}~\mathrm{eV}^2$, current best limit at
$3\sigma$ from the SK+K2K+MINOS combined analysis\cite{SKK2KMINOS}) and decay is not allowed in the tested model ($\alpha_3=0$).

\end{enumerate}

\begin{figure}[H]
    \centering
     \subfloat{
     \includegraphics[height=6cm]{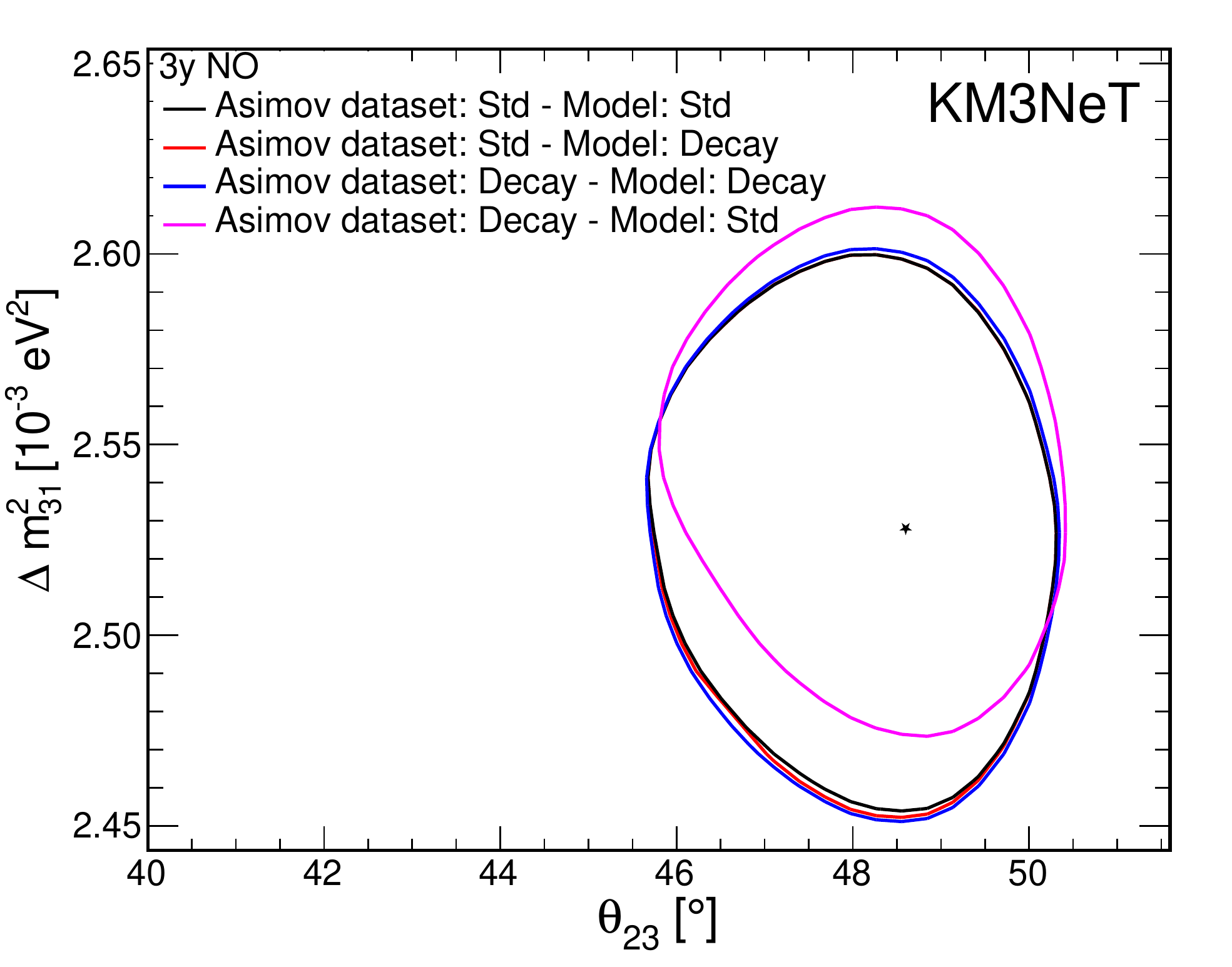}}
     \subfloat{
     \includegraphics[height=6cm]{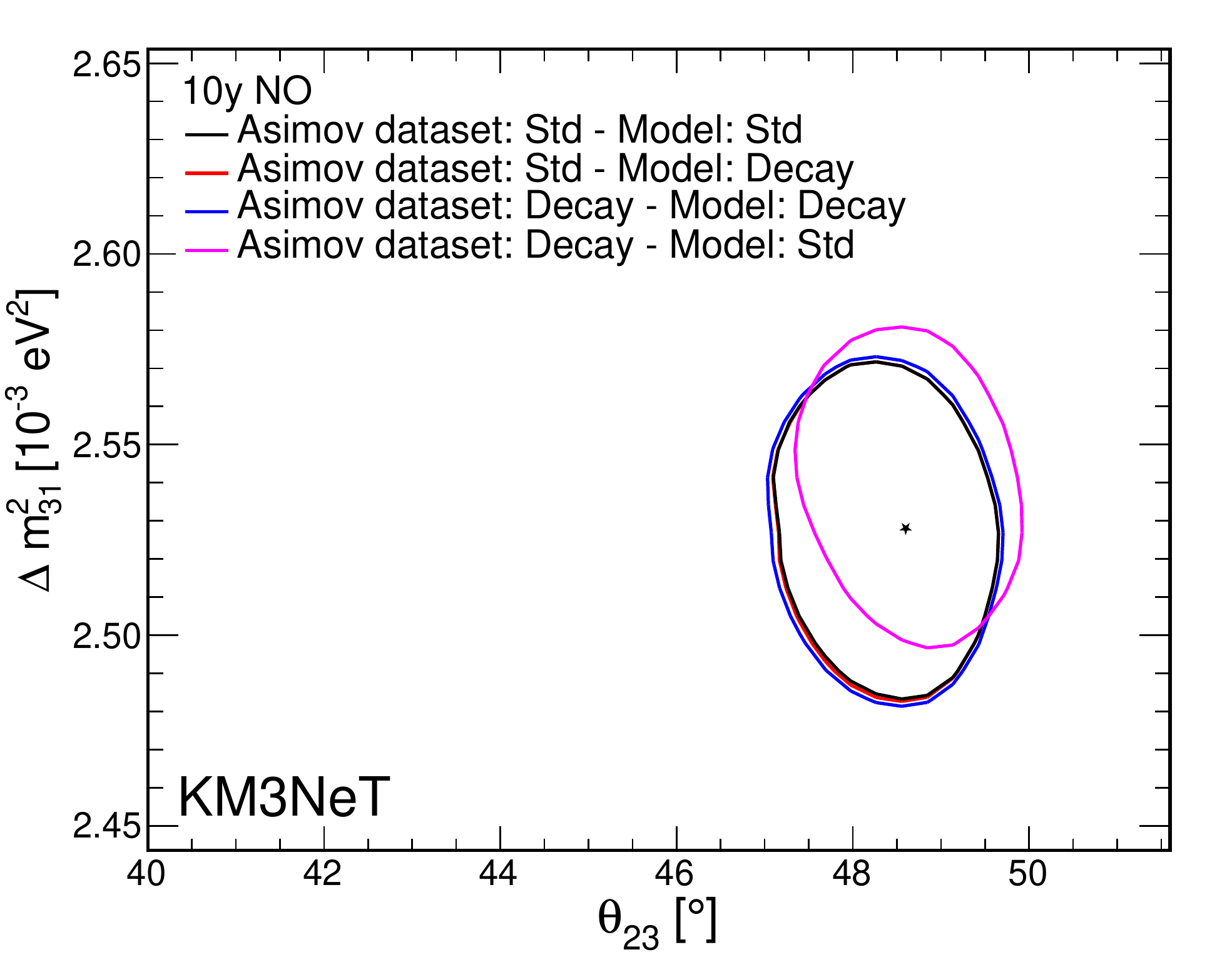}} \\
  \subfloat{
      \includegraphics[height=6cm]{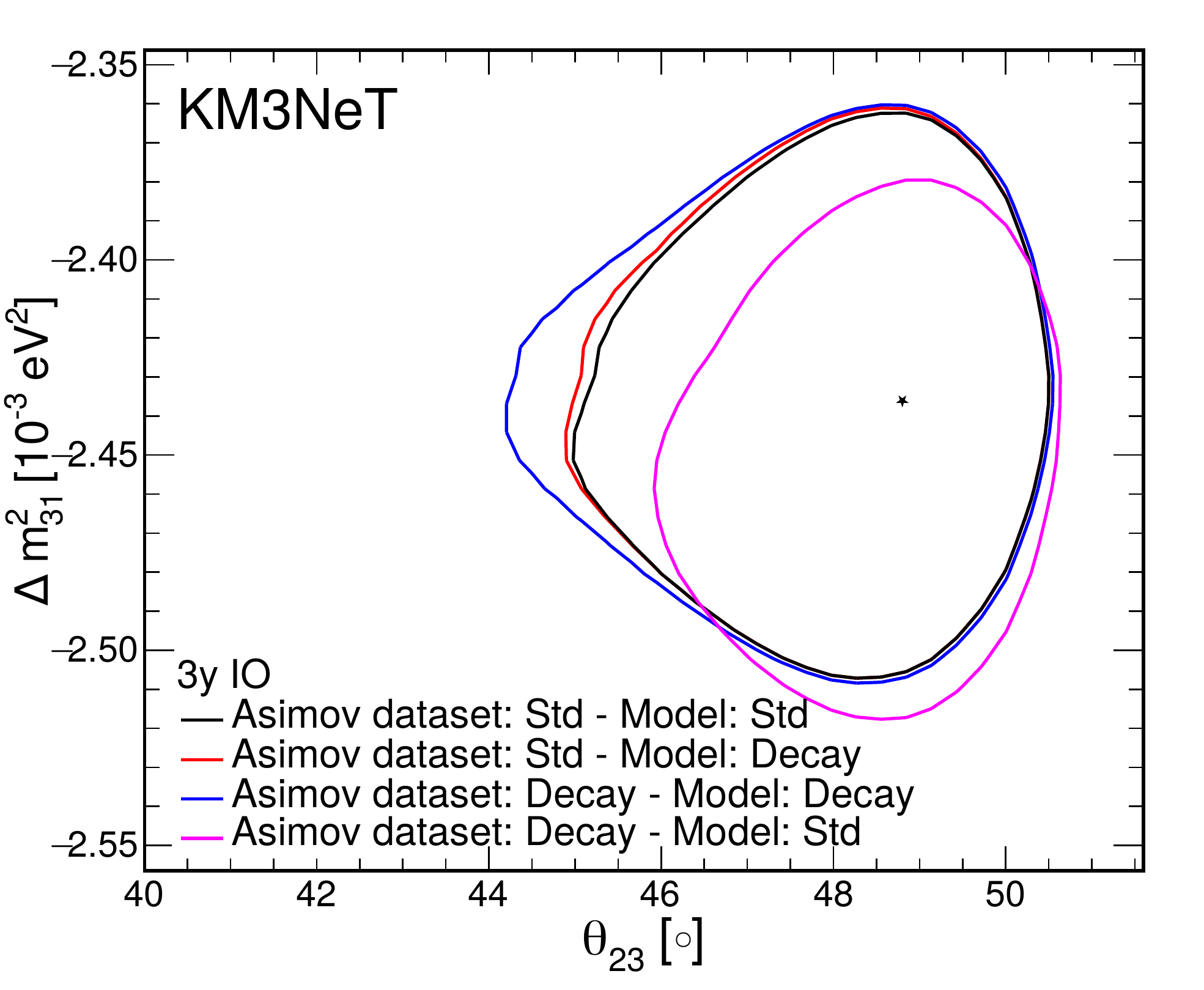}} 
  \subfloat{
      \includegraphics[height=6cm]{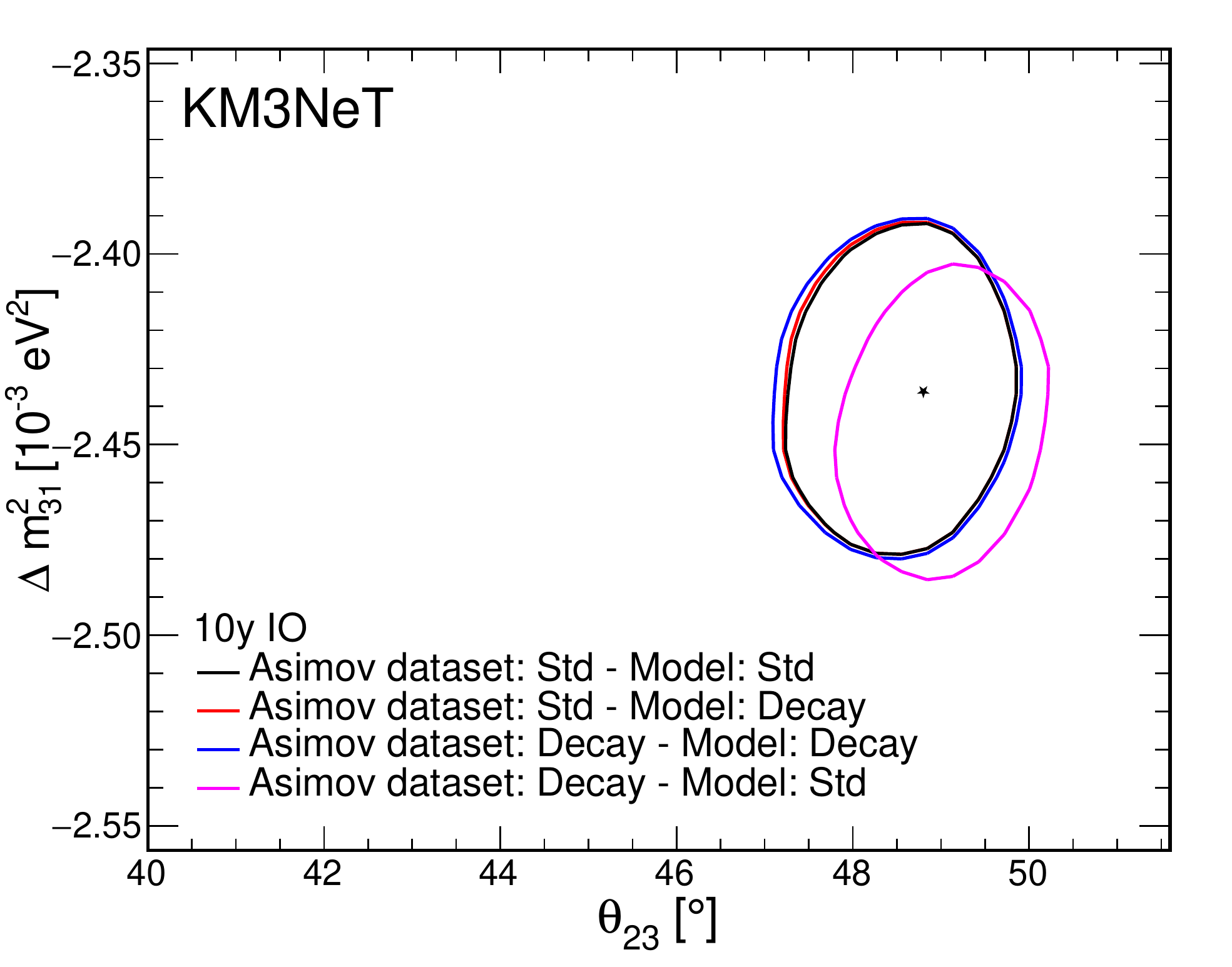}}    
      
    \caption{KM3NeT/ORCA 90\% CL sensitivity contour plots to the oscillation parameters $\theta_{23}$ and $\Delta m^2_{31}$ for 3 (left) and 10 (right) years of data taking assuming NO (top) or IO (bottom) for four cases: standard oscillations in both the Asimov dataset and model (black), standard oscillations in the Asimov dataset but using a decay model in the fit (red), Asimov dataset assuming a decay scenario ($\alpha_3=1.22\times 10^{-5}~\mathrm{eV}^2$) and decay model in the fit (blue), and  Asimov dataset assuming a decay scenario ($\alpha_3=1.22\times 10^{-5}~\mathrm{eV}^2$) but assuming standard oscillations in the model to fit (magenta). True values used to simulate the Asimov dataset are marked with stars.}
    \label{Fig12-ContourThetaDm}
\end{figure}

Figure \ref{Fig12-ContourThetaDm} presents the $90\%$ CL contour sensitivity plots for these four cases computed for 3 and 10 years of data taking and for both NMO hypotheses. There is almost no effect in allowing the model to fit the decay constant if neutrinos are stable (black and red) except for a slight increase of the width in the IO case  for small $\theta_{23}$ values. Assuming unstable neutrinos (blue) with the current $3\sigma$ limit for $\alpha_3$ claimed in ref. \cite{SKK2KMINOS}, the effects for the IO hypothesis get more pronounced. Finally, assuming unstable neutrinos (magenta) without allowing the decay model in the fit, the contours get deformed and shifted. This study shows that the precision with which KM3NeT will measure $\theta_{23}$ and $\Delta m^{2}_{31}$ will not be significantly affected by the presence of neutrino decay, but such a phenomenon may introduce biases in the measurement if it is not taken into account. This highlights the importance of testing BSM models to verify the robustness of the oscillation measurements.

\subsection{The impact of invisible neutrino decay on the NMO sensitivity.}

Mass ordering sensitivity is obtained by generating the Asimov dataset ($N^{\text{dat}}_{ij}$) assuming a true ordering, NO or IO, fitting both possible orderings in the model  ($N^{\text{mod}}_{ij}$) and getting the difference in the log-likelihood. NMO sensitivity in KM3NeT/ORCA depends on the true value of $\theta_{23}$, so a scan in the interval $[40^{\circ}, 50^{\circ}]$ is performed. This procedure is repeated for the four cases explained in section \ref{sec:Cases}.\footnote{
Note that in order to compute how neutrino decay affects the sensitivity to the NMO, the same framework \cite{Bourret:2018kug} is used in the four cases, which is slightly different from the one used in ref. \cite{NMOpaper}.}

\begin{figure}[H]
    \centering
     \subfloat{
     \includegraphics[height=6cm]{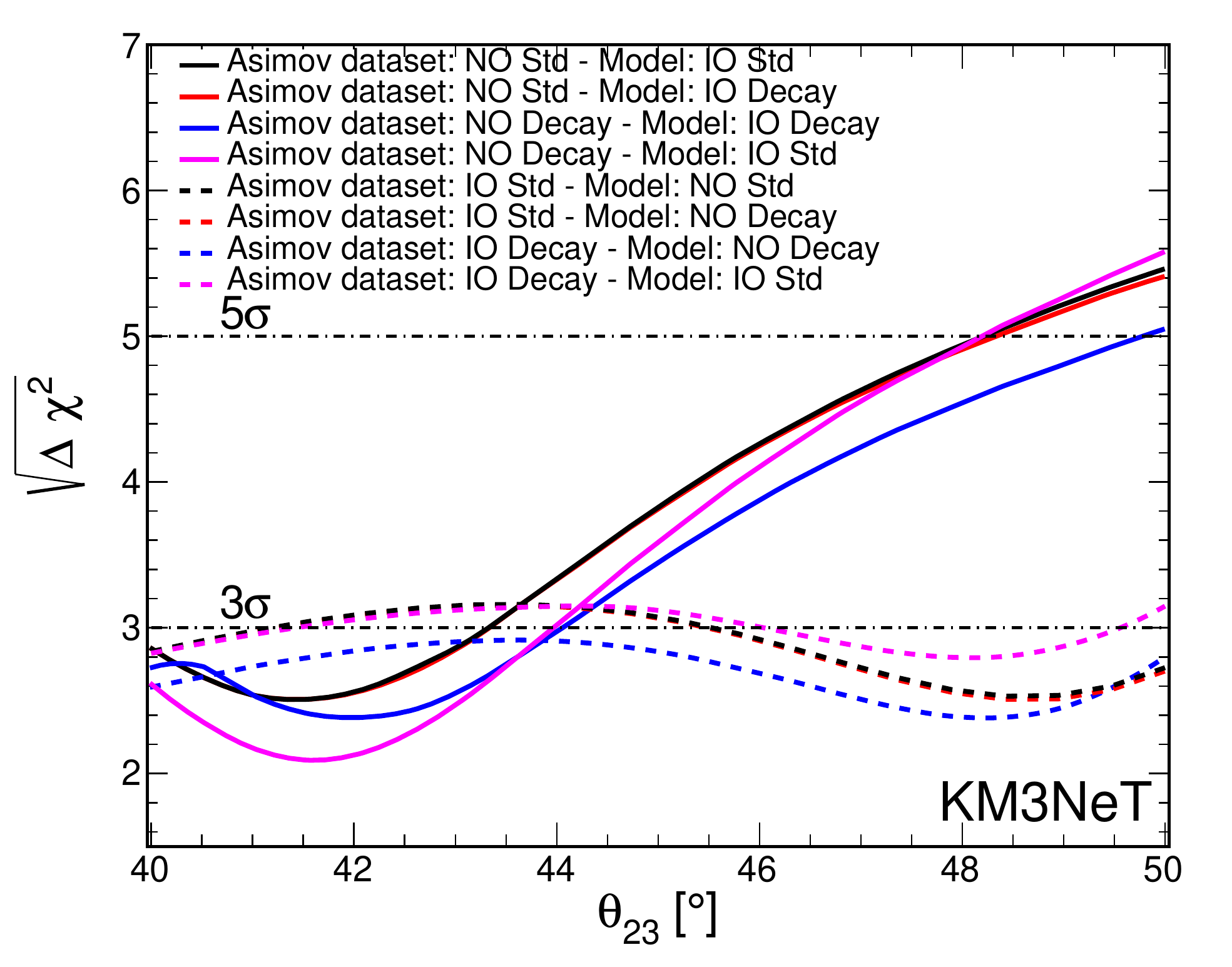}}
  \subfloat{
      \includegraphics[height=6cm]{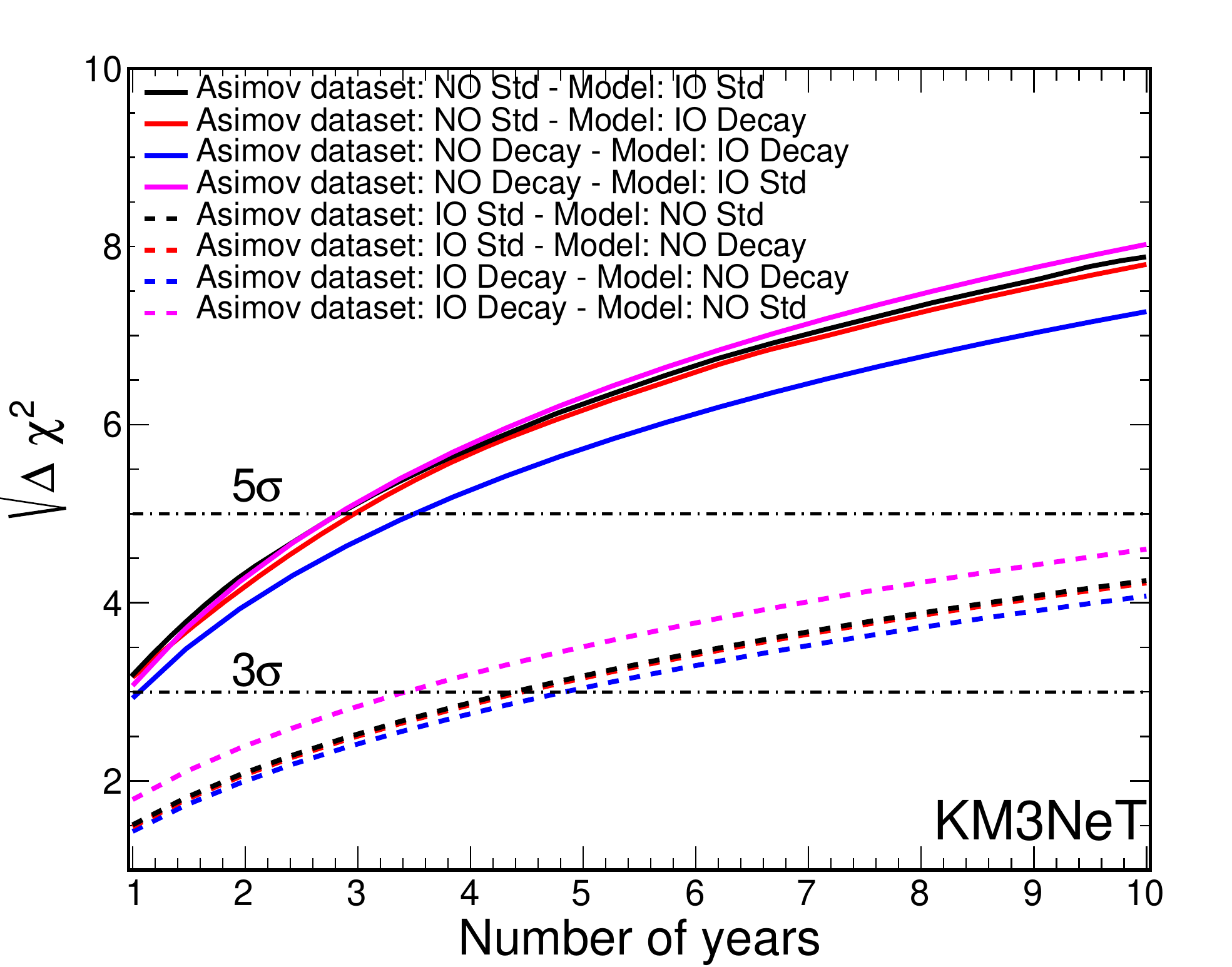}}
    \caption{Left: NMO sensitivity assuming true NO (solid) and true IO (dashed) for 3 years of data taking as a function of the true value of $\theta_{23}$. Right: NMO sensitivity assuming true NO (solid) and true IO (dashed) for $\theta_{23}=48.6^{\circ}$ as a function of the number of years of data taking. Four cases are considered: standard oscillations in both the Asimov dataset and the model (black), standard oscillations in the Asimov dataset and $\alpha_3$ free in the model (red), Asimov dataset assuming a decay scenario ($\alpha_3=1.22\times 10^{-5}~\mathrm{eV}^2$) and $\alpha_3$ free in the model (blue), and Asimov dataset assuming a decay scenario ($\alpha_3=1.22\times 10^{-5}~\mathrm{eV}^2$) but assuming standard
oscillations in the model (magenta). }
    \label{Fig13-MH}
\end{figure}

The sensitivity to the neutrino mass ordering as a function of the true value of $\theta_{23}$ is shown in figure \ref{Fig13-MH} (left) for 3 years of data taking.  Figure \ref{Fig13-MH} (right) shows the sensitivity as a function of the data taking time assuming as true $\theta_{23}$ the best-fit value according to NuFit 4.1 \cite{Esteban_2019} ($\theta_{23}=48.6^{\circ}$). Assuming that neutrinos are stable, there is almost no difference allowing the model to fit the decay constant (black and red curves), although the sensitivity is slightly reduced for large values of $\theta_{23}$. In the case of unstable neutrinos with a large decay value ($\alpha_3=1.22\times 10^{-5}~\mathrm{eV}^2$) when data is fitted to a decay model (blue), the mass ordering sensitivity decreases by $\sim 0.4\sigma$ ($0.2\sigma$) if NO (IO) is assumed over the whole range of $\theta_{23}$ except for $\theta_{23}=[40^{\circ},41^{\circ}]$ ($\theta_{23}>49.5^\circ$), where the sensitivity increases slightly above the standard oscillation case. However, if unstable neutrino data is fitted to a model assuming standard oscillations (magenta) the sensitivity to NMO would be slightly underestimated in the lower octant for NO and overestimated in the higher octant for IO. In the unstable neutrino scenario, there would be a small delay (around half a year) in determining which is the true ordering for the current best-fit from NuFit 4.1 ($\theta_{23}=48.6^{\circ})$.

\begin{figure}[H]
    \centering
    
     \includegraphics[height=8cm]{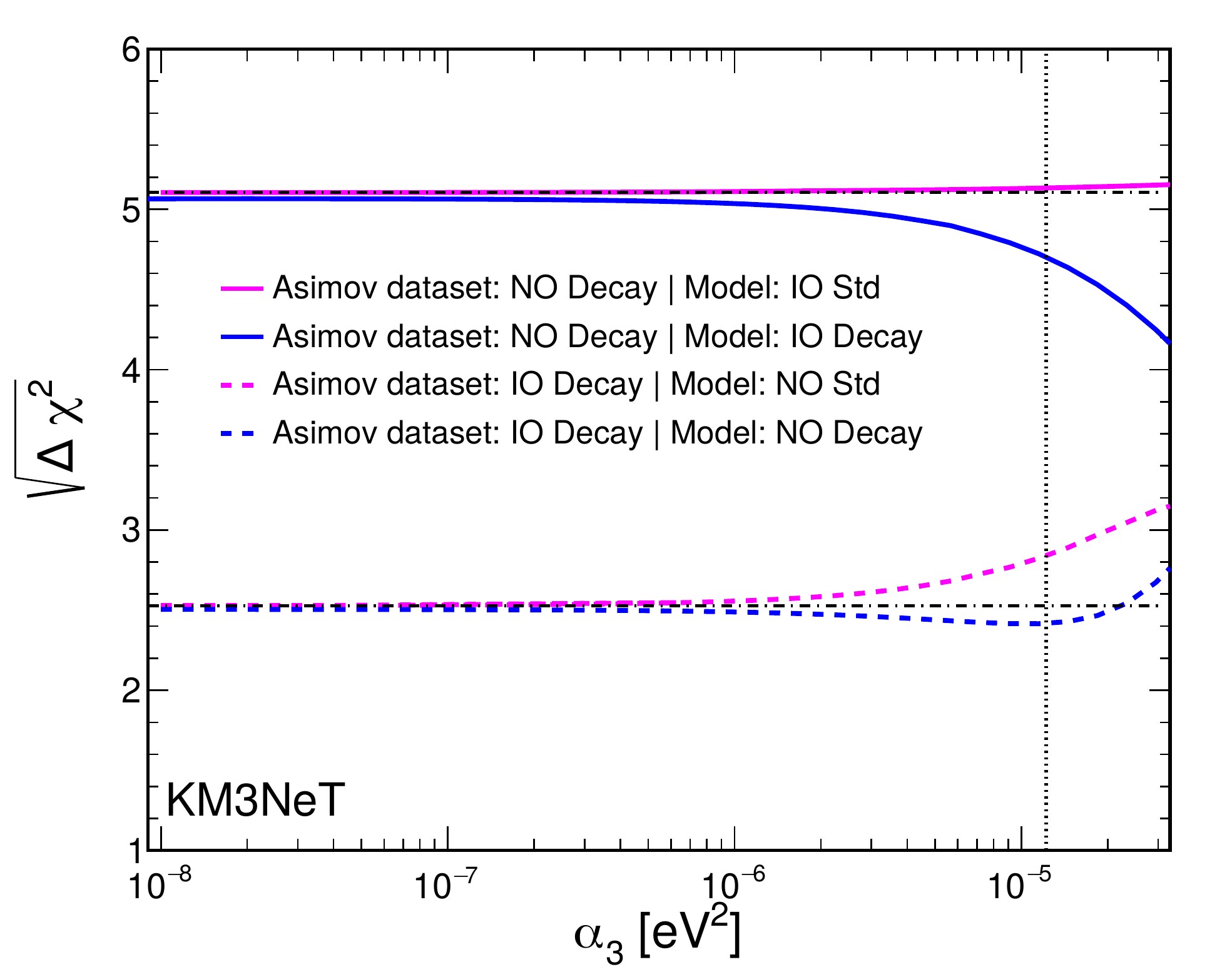}
    \caption{NMO sensitivity assuming true NO (solid) and true IO (dashed) for 3 years of data taking as a function of the true value of $\alpha_{3}$.  Two cases are considered: decay scenario in both the Asimov dataset and the model (blue) and decay scenario in the Asimov dataset, but assuming standard oscillations in the model (magenta). Horizontal lines correspond to the expected sensitivities when no invisible neutrino decay occurs. The vertical line corresponds to $\alpha_3=1.22\times 10^{-5}~\mathrm{eV}^2$ (current best limit from the combination of SK+K2K+MINOS at 3$\sigma$ CL \cite{SKK2KMINOS}).}
    \label{Fig14_NMOalpha}
\end{figure}

The NMO sensitivity as a function of the true $\alpha_3$ assumed in the Asimov dataset is displayed in figure \ref{Fig14_NMOalpha} for 3 years of data taking. The sensitivity is slightly affected up to $\alpha_3$ values close to the current $3\sigma$ limit \cite{SKK2KMINOS}, where the sensitivity is visibly reduced specially for true NO.

\section{Conclusions}

In this paper, the potential of KM3NeT/ORCA to search for the invisible neutrino decay of the neutrino mass eigenstate $\nu_3$ was probed. Upper limits at $90\%$ CL to the decay constant $\alpha_3$ have been computed for 3 and 10 years of data taking, corresponding to $5.7\times10^{-6}~\mathrm{eV}^2$ and  $3.7\times10^{-6}~\mathrm{eV}^2$, respectively. The lower limits in the inverse of $\alpha_3$ are $\tau_3/m_3=120~\mathrm{ps/eV}$ and $\tau_3/m_3=180~\mathrm{ps/eV}$, respectively. These results would improve current limits by up to two orders of magnitude. Depending on the true value of $\theta_{23}$ the lower limits on $\tau_3/m_3$ for 10 years of data taking move between $100$ ps/eV (the worst scenario, $\theta_{23}=40^{\circ}$) and $190$ ps/eV (the best scenario, $\theta_{23}=50^{\circ}$). Even in the worst scenario, KM3NeT/ORCA will be competitive with the current and future experiments.

 In the presence of unstable neutrinos, measurements with KM3NeT/ORCA would be precise enough to rule out the stable neutrino hypothesis with $3\sigma$ ($5\sigma$)  confidence if the true value of the decay constant is $\alpha_3>6.8 (12.0) \times 10^{-6}~\mathrm{eV^2}$ in 10 years of data taking and assuming NO.

The impact of neutrino decay on the precision for $\theta_{23}$ and $\Delta m^2_{31}$ measurements are almost negligible, slightly decreasing the KM3NeT/ORCA sensitivity to $\theta_{23}$. Neutrino decay effects in the neutrino mass ordering sensitivity were studied assuming unstable neutrinos as a function of the true value of $\theta_{23}$, revealing that, except for a very small interval, sensitivity can be reduced up to $\sim 0.4\sigma$ for NO and $\sim 0.2\sigma$ for IO compared with the standard oscillation scenario. Only large values of $\alpha_3$, currently strongly disfavoured by experimental results, could impact the sensitivity to NMO.

Events that contribute the most to the sensitivity to $\alpha_3$ are in the matter-resonant energy region, $3-8$ GeV. Future improvements in particle identification, selection efficiency and energy resolution at these energies would enhance significantly the sensitivity and future bounds.

\section{Acknowledgements}
The authors acknowledge the financial support of the funding agencies:
Agence Nationale de la Recherche (contract ANR-15-CE31-0020), Centre National de la Recherche Scientifique (CNRS), Commission Europ\'eenne (FEDER fund and Marie Curie Program), LabEx UnivEarthS (ANR-10-LABX-0023 and ANR-18-IDEX-0001), Paris \^Ile-de-France Region, France;
The General Secretariat of Research and Innovation (GSRI), Greece
Istituto Nazionale di Fisica Nucleare (INFN), Ministero dell'Universit\`a e della Ricerca (MIUR), PRIN 2017 program (Grant NAT-NET 2017W4HA7S) Italy;
Ministry of Higher Education, Scientific Research and Innovation, Morocco, and the Arab Fund for Economic and Social Development, Kuwait;
Nederlandse organisatie voor Wetenschappelijk Onderzoek (NWO), the Netherlands;
The National Science Centre, Poland (2021/41/N/ST2/01177);
National Authority for Scientific Research (ANCS), Romania;
Grants PID2021-124591NB-C41, -C42, -C43 funded by MCIN/AEI/ 10.13039/501100011033 and, as appropriate, by “ERDF A way of making Europe”, by the “European Union” or by the “European Union NextGenerationEU/PRTR”, Programa de Planes Complementarios I+D+I (refs. ASFAE/2022/023, ASFAE/2022/014), Programa Prometeo (PROMETEO/2020/019) and GenT (refs. CIDEGENT/2018/034, /2019/043, /2020/049. /2021/23) of the Generalitat Valenciana, Junta de Andaluc\'{i}a (ref. SOMM17/6104/UGR, P18-FR-5057), EU: MSC program (ref. 101025085), Programa Mar\'{i}a Zambrano (Spanish Ministry of Universities, funded by the European Union, NextGenerationEU), Spain;

\newpage
\bibliographystyle{JHEP}
\bibliography{main}
\end{document}